\ifdefined\pdfoutput\pdfoutput=1\fi
\documentclass[12pt]{article}
\usepackage{amssymb,amsmath,graphicx,xcolor,empheq,tikz-cd}
\usepackage[textwidth=17.5cm,textheight=23cm]{geometry}
\usepackage[status=draft,inline,nomargin]{fixme}

\makeatletter
\numberwithin{equation}{section}

\usepackage{graphicx}
\usepackage{dcolumn}
\usepackage{bm}
\usepackage{bbm}
\usepackage{framed}
\usepackage{setspace}
\usepackage{float}
\usepackage{mathtools}
\usepackage{stmaryrd}
\usepackage{authblk}
\usepackage{microtype}

\usepackage[pagebackref,linktoc=page]{hyperref}
\hypersetup{
colorlinks = true,
linkcolor= [rgb]{0.32, 0.38, 0.96},
citecolor=[rgb]{0.15,0.65,0.13},
urlcolor = [rgb]{0.25, 0.41, 0.68}}

\def\<{\langle}
\def\>{\rangle}

\DeclareTextFontCommand{\df}{\bf}

\usepackage{amsthm}

\newtheorem{prop}{Proposition}[section]
\newtheorem{conj}[prop]{Conjecture}
\newtheorem{cor}[prop]{Corollary}
\theoremstyle{definition}

\usetikzlibrary{arrows}

\def\[#1\]{%
  \begin{equation}\begin{gathered}#1\end{gathered}\end{equation}%
}

\newcommand{\Pic}{\underline{\underline{\mathsf{Pic}}}}
\newcommand{\BrPic}{\underline{\underline{\mathsf{BrPic}}}}
\newcommand{\U}{\mathrm{U}}
\newcommand{\PU}{\mathrm{PU}}
\newcommand{\SU}{\mathrm{SU}}
\newcommand{\SO}{\mathrm{SO}}
\newcommand{\rH}{\mathrm{H}}
\newlength{\simeqKwidth}

\title{Anomalies of TQFTs coupled to parameter spaces and constraints on ground-state degeneracies}
\author[1]{Takamasa Ando}
\author[2]{Ryan Thorngren}
\affil[1]{\normalsize\it Center for Gravitational Physics and Quantum Information,\protect\\
Yukawa Institute for Theoretical Physics, Kyoto University, Kyoto 606-8502, Japan}
\affil[2]{\normalsize\it Mani L. Bhaumik Institute for Theoretical Physics,\protect\\
Department of Physics and Astronomy, University of California, Los Angeles, CA 90095, USA}
\date{}

\begin{document}

\maketitle
\begin{abstract}
Anomalies of global symmetries can arise in long-range entangled quantum many-body systems and provide useful constraints on their dynamics.
In this paper, we study anomalies of $d$-dimensional topological quantum field theories (TQFTs) coupled to parameter spaces and their implications for ground-state degeneracies.
Specifically, we formulate anomalies as obstructions to constructing parametrized families of TQFTs.
We discuss how the classification of these anomalies depends on the choice of topology on the universal $\U(1)$ $(d-1)$-form symmetry.
In this framework, we use the projective ground-state bundle obtained by compactifying the TQFT on a spatial manifold to derive a general constraint relating its ground-state degeneracy $N$ to its torsion anomaly $\omega$, expressed as $N\int_W\omega\in\mathbb{Z}$ for bundles $W$ of spatial manifolds over closed surfaces.
We apply this constraint to ground-state degeneracies of $(2+1)$-dimensional TQFTs with framing and $1$-form symmetry anomalies and to obstructions to symmetry-preserving TQFTs. We also discuss implications for bordism generators and propose a conjecture on anomaly detection by manifolds fibered over surfaces.
\end{abstract}

%\clearpage
\tableofcontents
%\clearpage

% \section{Review of Kong-Wen}

\section{Introduction}
%TQFT
Topologically ordered systems are gapped quantum systems with long-range entanglement and have nontrivial excitations, such as particles, strings, and membranes~\cite{Wen:1989iv}. In $(2+1)$-dimensional topologically ordered systems, the particle excitations are called anyons, which have nontrivial braiding statistics~\cite{Kitaev:2005hzj}. 
The low-energy behavior of topologically ordered systems is described by topological quantum field theories (TQFTs). TQFTs are quantum field theories that are independent of local geometric data, such as the spacetime metric. TQFTs in general describe gapped quantum systems with or without intrinsic topological order, including topologically ordered systems and invertible phases.
A characteristic feature of intrinsic topological order is the presence of a nontrivial ground-state degeneracy that is robust against local perturbations. Such a degeneracy depends on the topology of the spatial manifold and is a topological invariant of the system~\cite{Wen:1990zza}.

%Anomaly
Ground states of intrinsically topologically ordered systems can neither be disentangled by local unitary transformations nor be realized as a unique gapped state on every closed spatial manifold.
Because of such long-range entanglement, the ground states of topologically ordered systems can carry anomalies of global symmetries, while the ground states of unique gapped Hamiltonians are believed to never carry nontrivial anomalies.
Anomalies of invertible global symmetries are formulated as obstructions to making the partition function of the system invariant under gauge transformations of the background gauge field. 
More specifically, for a finite symmetry group $G$, a background gauge field on spacetime $M^d$ is described by a classifying map $g\colon M^d\to BG$, with gauge transformations represented by homotopies.
Then the partition function of the system is said to be non-anomalous if it is invariant under homotopy of the map $g$, and anomalous otherwise. In particular, we often require that any ambiguity in the partition function be a $\U(1)$ phase controlled by the action of an invertible field theory in one higher dimension. More precisely, the partition function then takes values in the Hilbert space of this invertible theory, and is thus a section of a line bundle rather than a function. This is called the anomaly inflow mechanism~\cite{Callan:1984sa}.

%coupled to parameter spaces
In this paper, we study anomalies using this picture of theories coupled to parameter spaces. Such anomalies are also known as sigma model anomalies~\cite{Moore:1984dc,Moore:1984ws,Thorngren:2017vzn} or anomalies in spaces of coupling constants~\cite{Cordova:2019jnf,Cordova:2019uob}. In particular, we study anomalies of TQFTs coupled to parameter spaces, which are not necessarily classifying spaces of invertible symmetries.
We discuss how the anomalies are classified and how TQFTs can carry anomalies. To study general mathematical structures of anomalies, we consider the moduli space of TQFTs, which is a classifying space for parametrized families of TQFTs~\cite{Hsin:2020cgg,Aasen:2022cdu,Hsin:2022iug}.
There is a correspondence between TQFTs coupled to parameter spaces and parametrized families of TQFTs, supported by the Baez--Dolan--Lurie cobordism hypothesis~\cite{Baez:1995xq,Lurie:2009keu}.
We formulate anomalies of TQFTs coupled to parameter spaces as obstructions to constructing the corresponding parametrized families of TQFTs.
These obstructions are organized by the Postnikov tower of the moduli space of TQFTs, with the usual $\U(1)$-valued anomaly appearing at the final stage.
Here, the coefficient group $\U(1)$ arises from the universal $(d-1)$-form $\U(1)$ symmetry of the TQFT, which acts on quantum states by an overall phase. However, the choice of topology on this $\U(1)$ is sometimes important, for example, when considering continuous symmetries. We discuss two choices of topology on $\U(1)$: the standard topology and the discrete topology. The former corresponds to deformation classes of the bulk invertible theories, while the latter corresponds to isomorphism classes.\footnote{For instance, the $(1+1)$D invertible theories with partition functions $\exp(i\theta\int F/2\pi)$ are mutually non-isomorphic for different values of $\theta\in[0,2\pi)$, but they are all deformation equivalent. See, e.g.,~\cite{Freed:2016rqq,Yonekura:2018ufj}.}

After providing a general formulation of anomalies of TQFTs coupled to parameter spaces, we discuss the physical implications of such anomalies. 
Specifically, we consider a $d$-dimensional TQFT coupled to a parameter space $X$. Let $\Sigma^{d-1}$ be a spatial manifold and $N(\Sigma^{d-1},\hat{g})$ be the ground-state degeneracy of the system on $\Sigma^{d-1}$ with a parameter map $\hat{g}\colon\Sigma^{d-1}\to X$. 
One can think of this parameter map as a generalization of a background gauge field, which corresponds to the case $X=BG$ for a finite symmetry group $G$.
Let us denote a torsion anomaly of the system by $\omega$.
In this setup, Proposition~\ref{prop:anomaly-gsd} gives the general constraint:
\begin{equation}
    N(\Sigma^{d-1},\hat{g})\int_{W^{d+1}}g^*\omega\in\mathbb{Z},
\end{equation}
for any oriented closed $(d+1)$-dimensional manifold $W^{d+1}$ which fits into a fibration
\begin{equation}
    \Sigma^{d-1}\rightarrow W^{d+1}\rightarrow Y^2.
\end{equation}
Here, $Y^2$ is a closed connected surface, $g\colon W^{d+1}\to X$ is an arbitrary parameter map, and $\hat{g}$ is its restriction to a fiber.
We show this constraint by considering a compactification of the system on $\Sigma^{d-1}$, which gives a $(0+1)$-dimensional quantum mechanical system.
This constraint generalizes the constraint obtained by Kong and Wen for the framing anomaly~\cite{Kong:2020bmb}.
Based on this constraint, we also discuss several applications, such as constraints on the ground-state degeneracy of $(2+1)$-dimensional TQFTs with a nontrivial fractional quantum Hall response, and an obstruction to the existence of a symmetry-preserving TQFT.
We further discuss a conjecture on the detectability of anomalies by manifolds with $(d-1)$-dimensional fibers.

\paragraph{Organization of the paper}
This paper is organized as follows.
In Sec.~\ref{sec:anomaly_tqft_parameter}, we formulate anomalies using parametrized TQFTs and discuss their classification.
In Sec.~\ref{sec:anomaly_constraint}, we derive a general constraint on ground-state degeneracies, present its physical applications, and discuss anomaly detection by fibered manifolds.
We conclude with a discussion in Sec.~\ref{sec:discussion}.
Appendix~\ref{app:parametrized-tqft-details} provides further details on parametrized TQFTs.
Appendix~\ref{app:bulk-boundary-parametrized-tqft} discusses how bulk families act on boundary theories.
Appendix~\ref{app:anomaly_with_symmetry} studies obstructions to equipping parametrized families with a compatible global symmetry structure.
Readers primarily interested in the ground-state degeneracy constraints and their applications may proceed directly to Sec.~\ref{sec:anomaly_constraint}.

\paragraph{Notation}
$B^{p+1}G$ denotes the classifying space of a $p$-form symmetry $G$. For finite $G$, it is homotopy equivalent to the Eilenberg--MacLane space $K(G,p+1)$.
For a space $X$ with a $G$-action, we write $X\mathord{\sslash}G=(X\times EG)/G$ for its homotopy quotient, where $EG$ is a contractible space with a free $G$-action and the quotient is taken by the diagonal $G$-action.
$\Omega_n^{S}(X)$ denotes the bordism group of closed $n$-manifolds equipped with a tangential structure $S$ and a map to a parameter space $X$.
We write the shifted Anderson dual as $\Omega^{n}_{S}(X)\coloneqq (I_{\mathbb{Z}}\Omega^{S})^{n+1}(X)$, which classifies deformation classes of reflection-positive invertible extended field theories in $n$ spacetime dimensions with the same background structures~\cite{Freed:2016rqq,Grady:2023sav}.
We use superscripts to indicate manifold dimensions; in particular, $M^n$ denotes an $n$-dimensional spacetime manifold.

\section{Anomalies as obstructions to parametrized TQFTs}\label{sec:anomaly_tqft_parameter}
In this section, we formulate anomalies of TQFTs coupled to parameter spaces as obstructions to the existence of parametrized families of TQFTs.
We first review parametrized TQFTs and their moduli spaces,
and then discuss how the topology assigned to the universal $\U(1)$ $(d-1)$-form symmetry affects the anomaly classification.
We illustrate this framework with the fractional quantum Hall response and the framing anomaly, and discuss its dynamical consequences.

\subsection{TQFTs coupled to parameter spaces}\label{sec:tqft_parameter}
In this paper we consider anomalies of QFTs coupled to parameter spaces.
To explain such anomalies, we consider a theory in $d$ spacetime dimensions on a manifold $M^d$ equipped with a continuous map $g\colon M^d\to X$ to a parameter space $X$.
We say that the theory is non-anomalous if the evaluation of the theory, say the partition function, only depends on the homotopy class of the map $g$, and anomalous otherwise.

This definition generalizes the usual anomaly of a finite global $G$ symmetry, where the parameter space is given by the classifying space $X=BG$ of the group $G$.
If the theory has an 't Hooft anomaly, then the partition function is not invariant under homotopies of the map $g$, which represent gauge transformations.
We often further require that any ambiguity in the (normalized) partition function be a $\U(1)$ phase controlled by the action of an invertible field theory in one higher dimension.
For a continuous symmetry $G$, one also needs to introduce geometric data, such as a connection on the $G$-bundle, to define the partition function in the presence of a background gauge field.
Then two background gauge fields whose underlying $G$-bundles are isomorphic are not necessarily gauge equivalent, and the partition function may depend on the connection even when it is gauge invariant.
Nevertheless, using the parameter space $X=BG$ remains useful for studying continuous symmetries.
In Sec.~\ref{sec:anomaly_parameter}, we will provide a definition of anomalies of parametrized TQFTs, which gives the same classification of anomalies for the classifying space of a finite group.
Although this formulation is not necessarily equivalent to the usual definition of global anomalies for continuous symmetries, deformation classes of invertible theories admit a description in terms of topological data~\cite{Freed:2016rqq,Yamashita:2021cao,Grady:2023sav}.
We discuss anomalies for continuous group symmetries in detail in Sec.~\ref{sec:conti_sym}.

A main focus of this paper is on anomalies of TQFTs, and we formulate an anomaly of a $d$-dimensional TQFT coupled to parameter spaces in terms of anomalies of $X$-parametrized TQFTs.
To proceed, let us recall the Atiyah--Segal formulation of TQFTs~\cite{Atiyah:1989vu,Segal:2002ei}, which states that a $d$-dimensional TQFT is a symmetric monoidal functor from the category of $d$-dimensional bordisms to the category of vector spaces:
\begin{equation}
    Z\colon\mathrm{Bord}_{d}^{S}(X) \rightarrow \mathrm{Vect}_{\mathbb{C}}.
\end{equation}
Here, $\mathrm{Bord}_{d}^{S}(X)$ is the category of $d$-dimensional bordisms with a structure $S$ and a map to a parameter space $X$, and $\mathrm{Vect}_{\mathbb{C}}$ is the category of complex vector spaces.
Physically, the functor $Z$ is the partition function of the TQFT, which assigns a complex number to a closed $d$-dimensional spacetime manifold, and a Hilbert space to a closed $(d-1)$-dimensional spatial manifold.
The formulation of such functorial TQFTs is generalized to extended TQFTs, where we also have additional assignments to lower-dimensional manifolds.

The Baez--Dolan--Lurie cobordism hypothesis~\cite{Baez:1995xq,Lurie:2009keu} states that fully extended framed TQFTs are classified by fully dualizable objects in some target higher category.
\begin{equation}
    \operatorname{Fun}^\otimes(\mathrm{Bord}_d^{\mathrm{fr}}, \mathcal{C}) \simeq \mathcal{C}^{\mathrm{fd}}.
\end{equation}
Here, $\mathrm{Bord}_d^{\mathrm{fr}}$ is the $(\infty,d)$-category of $d$-dimensional framed bordisms, $\mathcal{C}$ is a symmetric monoidal $(\infty,d)$-category, and $\mathcal{C}^{\mathrm{fd}}$ is the maximal $\infty$-groupoid of fully dualizable objects in $\mathcal{C}$.
The symbol $\simeq$ denotes an equivalence of $\infty$-groupoids, or equivalently, a weak homotopy equivalence of the corresponding spaces.
For a tangential structure $S\rightarrow \mathrm{O}(d)$ and a decoupled parameter space $X$, we also have the following 
parametrized version of the cobordism hypothesis~\cite[Thms.~2.4.18 and~2.4.26]{Lurie:2009keu}:
\begin{equation}
    \operatorname{Fun}^\otimes(\mathrm{Bord}_d^{S}(X), \mathcal{C}) \simeq \operatorname{Map}(X, (\mathcal{C}^{\mathrm{fd}})^{hS}).
\end{equation}
Here, $(\mathcal{C}^{\mathrm{fd}})^{hS} \coloneqq \operatorname{Map}_{S}(ES,\mathcal{C}^{\mathrm{fd}})$ is the homotopy fixed point space,\footnote{Here $\operatorname{Map}_S(A,B)$ denotes the space of continuous $S$-equivariant maps, satisfying $f(s\cdot a)=s\cdot f(a)$.} where $S$ acts through $S\to \mathrm{O}(d)$ and the $\mathrm{O}(d)$ action induced by changes of framing.
To understand this statement physically, note that the set of connected components of the right-hand side is given by
\begin{equation}
    \pi_0\operatorname{Map}(X, (\mathcal{C}^{\mathrm{fd}})^{hS})\cong [X,(\mathcal{C}^{\mathrm{fd}})^{hS}]
    =\operatorname{Map}(X, (\mathcal{C}^{\mathrm{fd}})^{hS})/\text{homotopy}.
\end{equation}
Now this classifies parametrized families of TQFTs over $X$ as we explain below.
As discussed in~\cite[Sec.~VII]{Hsin:2020cgg}, the cobordism hypothesis gives a correspondence between TQFTs coupled to parameter spaces and parametrized families of TQFTs.
Using this correspondence, we will define an anomaly of TQFTs coupled to parameter spaces as an obstruction to having corresponding parametrized TQFTs in Sec.~\ref{sec:anomaly_parameter}.

\subsection{Moduli space of parametrized TQFTs}\label{sec:moduli_tqft}
In this paper, we mainly consider \textit{simple} TQFTs, which are TQFTs with a unique ground state on the sphere $S^{d-1}$.
The moduli space $\mathcal{M}$ of a topologically ordered system is defined to be a classifying space of parametrized families of TQFTs.
Namely, there is a correspondence between
\begin{equation}
    \{\text{Parametrized families of a TQFT over }X\}/\text{continuous deformations}
\end{equation}
and
\begin{equation}
    [X,\mathcal{M}]\coloneqq \operatorname{Map}(X, \mathcal{M})/\text{homotopy}.
\end{equation}
The moduli space $\mathcal{M}$ encodes the higher categorical structure of invertible topological symmetry operators in the system.
For instance, the moduli space of a $(2+1)$D TQFT specified by a modular tensor category $\mathcal{B}$ is conjectured to be $B\Pic(\mathcal{B})$, where $\Pic(\mathcal{B})$ is the Picard 2-group of invertible $\mathcal{B}$-module categories~\cite{Kitaev:2005hzj,Aasen:2022cdu,Hsin:2022iug}.%
\footnote{In addition to a modular tensor category, one also needs to specify the chiral central charge to define a $(2+1)$D TQFT; for bosonic theories, the category determines this charge only modulo $8$. More precisely, we consider the moduli space with this additional datum fixed, yet $B\Pic(\mathcal{B})$ is still expected to capture all invertible topological defects of the theory.}

Physically, the $n$-th homotopy group $\pi_n\mathcal{M}$ of the moduli space $\mathcal{M}$ is interpreted as a group of topological defects of codimension $n$ in spacetime, namely the group of invertible $(n-1)$-form symmetries.
Thus we naturally expect the moduli space of a $d$-dimensional TQFT to be a homotopy $(d+1)$-type, i.e., $\pi_{>d+1}\mathcal M=0$.
For any $d$-dimensional simple TQFT, we have a universal $(d-1)$-form $\U(1)$ symmetry associated with the freedom to multiply quantum states by an overall phase.

To be more specific, let us first see the moduli space of $(0+1)$D simple TQFTs, which are quantum mechanical systems with a unique ground state. The moduli space of $(0+1)$D simple TQFTs is given by $B\U(1)\simeq K(\mathbb{Z},2)$, which is the classifying space for complex line bundles.

To understand the homotopy type of the moduli space $\mathcal{M}$ in higher dimensions, it is useful to consider a Postnikov tower of $\mathcal{M}$. The Postnikov tower of a connected CW complex $X$ is a sequence of spaces $X_{n}$, which are $n$-truncated, i.e., $\pi_{>n}X_{n}=0$, with compatible maps $X\to X_n$ inducing isomorphisms on $\pi_i$ for $i\leq n$:
\begin{equation}
    X\rightarrow \cdots \rightarrow X_{n+1} \xrightarrow{p_{n+1}} X_{n} \xrightarrow{p_n} X_{n-1} \rightarrow \cdots \rightarrow X_{0}.
\end{equation}
For $n\geq1$, the homotopy fiber of $p_{n+1}$ is $K(\pi_{n+1}X,n+1)$, and the fibration is classified by a \textit{Postnikov class} $\omega_{n}\in \rH^{n+2}(X_n,\pi_{n+1}X)$, with local coefficients determined by the action of $\pi_1X$ on $\pi_{n+1}X$.
Every connected CW complex admits such a tower, unique up to homotopy equivalence compatible with the tower maps and the maps from $X$.
We give a more detailed discussion of parametrized families of TQFTs using the Postnikov tower of the moduli space in Appendix~\ref{app:parametrized-tqft-details}.

\subsection{Anomaly of parametrized TQFTs}\label{sec:anomaly_parameter}
Let us now define an anomaly of a TQFT coupled to parameter spaces as an anomaly of the corresponding parametrized TQFT.
Consider a $d$-dimensional TQFT and a (connected) parameter space $X$. An anomaly of an $X$-parametrized TQFT is defined to be an obstruction to the existence of a genuine parametrized family of TQFTs over $X$.
More specifically, we define a $k$-th anomaly of $X$-parametrized TQFTs to be an obstruction to the existence of a lift of the map $f_{k}\colon X\rightarrow \mathcal{M}_{k}$ to the $(k+1)$-th Postnikov stage $\mathcal{M}_{k+1}$ of the moduli space $\mathcal{M}$. The $k$-th anomaly is classified by a cohomology class $f_{k}^*\omega_{k}\in \rH^{k+2}(X,\pi_{k+1}\mathcal{M})$, where $\omega_{k}$ is the Postnikov class of the fibration $p_{k+1}\colon \mathcal{M}_{k+1}\rightarrow \mathcal{M}_{k}$. Here the coefficient group is in general twisted by the action of $\pi_1 X\rightarrow\pi_1\mathcal{M}$ on $\pi_{k+1}\mathcal{M}$.
When the action is trivial, the lifting problem can be represented by the following diagram:
\begin{equation}
    \begin{tikzcd}
        K(\pi_{k+1}\mathcal{M},k+1) 
            \arrow[r]
        &
        \mathcal{M}_{k+1}
            \arrow[d,"p_{k+1}"]
            \arrow[r]
        & 
        * 
            \arrow[d]
        \\
        X
            \arrow[ru, dashed, "\exists\,f_{k+1}"{description}]
            \arrow[r,"f_{k}"']
        &
        \mathcal{M}_{k} 
            \arrow[r,"\omega_{k}"']
        &
        K(\pi_{k+1}\mathcal{M},k+2)\mathrlap{.}
    \end{tikzcd}
\end{equation}
If the $k$-th anomaly $f_{k}^*\omega_{k}$ vanishes, then lifts of $f_{k}$ correspond to trivializations of $f_{k}^*\omega_{k}$, and their homotopy classes form a torsor for $\rH^{k+1}(X,\pi_{k+1}\mathcal{M})$.
Further obstruction classes can depend on the choice of trivialization.
We note that the $k$-th anomaly here is different from the usual 't Hooft anomaly.
For instance, consider the case where the parameter space $X$ is a classifying space $BG$ of a finite group $G$. 
Then a $k$-th anomaly is an obstruction to having a symmetry-enriched topological (SET) model with symmetry $G$~\cite{Barkeshli:2014cna}. In $(2+1)$ dimensions, the $1$-st anomaly takes values in $\rH^3(BG,\mathcal{A}_\rho)$, where $\mathcal{A}=\pi_2\mathcal{M}$ is the group of abelian anyons and $\rho$ is the induced $G$ action. This is the $\rH^3$-obstruction in the literature. If this obstruction is nontrivial, the $G$ symmetry is actually part of a $2$-group symmetry, rather than just a $0$-form $G$ symmetry~\cite{Thorngren:2015gtw,Tachikawa:2017gyf,Benini:2018reh}. If this obstruction vanishes, its trivializations form a torsor for $\rH^2(BG,\mathcal{A}_\rho)$ and are called symmetry fractionalization classes.
The usual 't Hooft anomaly appears as an obstruction to having a lift to the top of the Postnikov tower of the moduli space $\mathcal{M}$, which is classified by a cohomology class in $\rH^{d+1}(BG,\U(1))$.
We refer to the obstruction associated with the universal scalar $\U(1)$ as the $\U(1)$-valued anomaly, or simply the anomaly.
In quantum lattice systems, anomaly indices have also been identified with pullbacks of Postnikov classes of higher symmetry groups~\cite{Kapustin:2025nju}; see also~\cite{Else:2014vma,Kawagoe:2025ldx,Czajka:2025mme} for related discussions.

However, when considering continuous group symmetries, or general parameter spaces, the topology on the universal $(d-1)$-form $\U(1)$ symmetry becomes important.
In the following, we discuss two choices of topology on $\U(1)$: the standard topology and the discrete topology. 
The standard topology is the usual topology on $\U(1)$ as a Lie group $\U(1)\simeq K(\mathbb{Z},1)$, while the discrete topology is the topology on $\U(1)$ as a discrete group.
The former corresponds to the deformation class of the anomaly, while the latter corresponds to the isomorphism class of the anomaly.
We further note that the natural inclusion $\U(1)\hookrightarrow\mathbb{C}^{\times}$ is a homotopy equivalence for the standard topologies, so the two choices give the same deformation classes. This inclusion is not a homotopy equivalence for the discrete topologies. Throughout this paper, we take the universal scalar group to be $\U(1)$, identifying scalar automorphisms that differ by a positive real factor in the moduli spaces under consideration.
This convention is applied separately to each vacuum sector.

\paragraph*{Deformation class}
When we take the standard topology on $\U(1)$, the Postnikov tower of the moduli space $\mathcal{M}$ is given by
\begin{equation}
    \begin{tikzcd}
        K(\mathbb{Z},d+1)
            \arrow[r]
        &
        \mathcal{M}
            \arrow[d]
        \\
        K(\pi_{d-1}\mathcal{M},d-1)
            \arrow[r]
        &
        \mathcal{M}_{d-1}
            \arrow[d]
        \\
        &
        \vdots \, 
    \end{tikzcd}
\end{equation}

The fiber $K(\mathbb{Z},d+1)$ encodes the higher Berry class~\cite{Kapustin:2020eby,Hsin:2020cgg,Wen:2021gwc,Kapustin:2022apy,Choi:2022odr,Ohyama:2023suc,Artymowicz:2023erv,Qi:2023ysw,Spiegel:2025ugu}.
For $d=1$, a map $X\to K(\mathbb{Z},2)\simeq B\U(1)$ determines a complex line bundle over $X$, classified up to isomorphism by its first Chern class in $\rH^2(X,\mathbb{Z})$.
More generally, a chosen lift of a family map $f\colon X\to\mathcal{M}$ to this fiber defines an integral higher Berry class through $[X,K(\mathbb{Z},d+1)]\cong\rH^{d+1}(X,\mathbb{Z})$, generalizing the first Chern class of the Berry line bundle.
In $(2+1)$ dimensions, the fiber is $K(\mathbb{Z},4)$, and the higher Berry class lies in $\rH^4(X,\mathbb{Z})$.
The obstruction to such a lift is discussed in Appendix~\ref{app:berry-lift}.

\paragraph*{Isomorphism class}
On the other hand, when we take the discrete topology on $\U(1)$, the Postnikov tower of the moduli space $\mathcal{M}$ looks like
\begin{equation}
    \begin{tikzcd}
        K(\U(1)_{\mathrm{disc}},d)
            \arrow[r]
        &
        \mathcal{M}
            \arrow[d]
        \\
        K(\pi_{d-1}\mathcal{M},d-1)
            \arrow[r]
        &
        \mathcal{M}_{d-1}
            \arrow[d]
        \\
        &
        \vdots
    \end{tikzcd}
\end{equation}
We explain the physical meaning of the two different choices of topology on $\U(1)$ in the following examples.

\paragraph*{Remark on non-simple TQFTs}
In non-simple TQFTs whose ground-state degeneracy on the sphere $S^{d-1}$ arises from a decomposition into simple sectors, we also have other $(d-1)$-form $\U(1)$ symmetries corresponding to the $\mathrm{U}(1)$ phase degree of freedom associated with each degenerate ground state.
The universal $(d-1)$-form $\U(1)$ symmetry is the diagonal subgroup of such $(d-1)$-form $\U(1)$ symmetries.
More generally, we define the projective moduli space $P\mathcal{M}$ by projectivizing with respect to the universal scalar $\U(1)$.
Then the $\U(1)$-valued anomaly is defined to be an obstruction to having a lift of the map $X\rightarrow P\mathcal{M}$ to $\mathcal{M}$.
This lifting problem need not coincide with a stage of the Postnikov tower.

\paragraph{Framing anomaly}
For a modular tensor category $\mathcal{B}$, one can construct a $(2+1)$D TQFT called the Reshetikhin--Turaev (RT) TQFT~\cite{Reshetikhin:1991tc,Turaev:1994xb}.
To define the RT theory functorially on bordisms, i.e., in a way consistent with cutting and gluing, one generally needs to choose an additional tangential structure, such as a $2$-framing or an oriented $p_1$-structure~\cite{Freed:2026fcn}.
The theory so defined depends on this choice: for instance, if one implements a unit shift of the $p_1$-structure, the partition function of the RT theory on a closed oriented $3$-manifold $M^3$ acquires a phase factor $\exp(2\pi i c_-/24)$, where $c_-$ is the chiral central charge of the theory~\cite{Witten:1988hf,FreedScheimbauerTeleman2026}.
When this phase is nontrivial, the theory does not descend to an oriented TQFT, and this obstruction is called the framing anomaly.

We explain that this obstruction is understood as an anomaly for the choice of the discrete topology on $\U(1)$.
To see this, note that the cobordism hypothesis for $(2+1)$D oriented TQFTs gives the following correspondence~\cite[Thm.~2.4.26]{Lurie:2009keu}:
\begin{equation}
    \operatorname{Fun}^\otimes(\mathrm{Bord}_3^{\SO(3)},\mathcal{C})\simeq (\mathcal{C}^{\mathrm{fd}})^{h\SO(3)},
\end{equation}
and the RHS is equivalent to
\begin{equation}
    (\mathcal{C}^{\mathrm{fd}})^{h\SO(3)}\simeq \Gamma(B\SO(3),\mathcal{C}^{\mathrm{fd}}\mathord{\sslash} \SO(3)\rightarrow B\SO(3)).
\end{equation}
Here, $\Gamma(B,E\to B)$ denotes the space of sections of the map $E\to B$.
We may use $B\Pic(\mathcal{B})$ if the TQFT is specified by a modular tensor category $\mathcal{B}$.
Note that $B\Pic(\mathcal{B})\mathord{\sslash} \SO(3)=(B\Pic(\mathcal{B})\times E\SO(3))/\SO(3)$ fits into the following fibration
\begin{equation}
    B\Pic(\mathcal{B})\rightarrow B\Pic(\mathcal{B})\mathord{\sslash} \SO(3)\rightarrow B\SO(3),
\end{equation}
and then consider the fiberwise Postnikov tower of the fibration. Namely, the $n$-th stage of the tower $\mathcal{M}_n$ looks like 
\begin{equation}
    \pi_{\leq n}B\Pic(\mathcal{B})\rightarrow \mathcal{M}_n\rightarrow B\SO(3),
\end{equation}
and the projection $p_{n+1}\colon \mathcal{M}_{n+1}\rightarrow \mathcal{M}_n$ fits into 
\begin{equation}
    \begin{tikzcd}
        K(\pi_{n+1}B\Pic(\mathcal{B}),n+1)
            \arrow[r]
        &
        \mathcal{M}_{n+1}
            \arrow[d,"p_{n+1}"]
            \arrow[r]
        &
        *
            \arrow[d]
        \\
        &
        \mathcal{M}_n
            \arrow[r,"\omega"']
        &
        K(\pi_{n+1}B\Pic(\mathcal{B}),n+2)\mathrlap{.}
    \end{tikzcd}
\end{equation}
Here the diagram is written assuming that the action of $\pi_1(\mathcal{M}_n)$ on $\pi_{n+1}B\Pic(\mathcal{B})$ is trivial.
In general, the $k$-invariant is a cohomology class with local coefficients, but the coefficient system becomes constant after pullback to $B\SO(3)$, since $B\SO(3)$ is simply connected.
Suppose we have a section $g_2\colon B\SO(3)\rightarrow \mathcal{M}_2$. Then the obstruction to lifting $g_2$ to $\mathcal{M}_3$ is given by a cohomology class $g_2^*\omega\in \rH^4(B\SO(3),\pi_3B\Pic(\mathcal{B}))$.
Since we are now taking the discrete topology on $\U(1)$, we have $\pi_3B\Pic(\mathcal{B})\cong \U(1)_{\mathrm{disc}}$, and so the obstruction class is an element of $\rH^4(B\SO(3),\U(1)_{\mathrm{disc}})$.
For a $(2+1)$D TQFT with a chiral central charge $c_-$, the obstruction class is given by $(c_-/24)p_1$.
In Sec.~\ref{sec:framing_anomaly}, we further discuss this framing anomaly from the perspective of the metric dependent local counterterm, known as the gravitational Chern--Simons term.

\subsection{Continuous symmetries}\label{sec:conti_sym}
So far, we have considered parameter spaces described purely as topological spaces, such as classifying spaces of finite groups.
When coupling to a background gauge field for a continuous symmetry, we also need geometric data, such as a connection, in addition to a classifying map $M^d\rightarrow BG$.
To describe coupling to a background gauge field for a continuous symmetry $G$, it is convenient to use $B_{\nabla}G$, the classifying stack of principal $G$-bundles with connection, which carries a universal principal $G$-bundle with connection~\cite[Def.~3.2.4 and Prop.~3.2.5]{Fiorenza:2012cech}.
The background gauge field is then obtained by a classifying map
\begin{equation}
    g_{\nabla}\colon M^d\rightarrow B_{\nabla}G.
\end{equation}
If one cannot make the partition function $Z(M^d,g_{\nabla})$ invariant under gauge transformations, we say that the theory has an 't Hooft anomaly of the $G$ symmetry.

Let us consider a system with a non-anomalous symmetry $G$ and ask whether we can get rid of the connection dependence of the partition function by adding a local counterterm. Namely, we ask whether the corresponding parametrized TQFT can be lifted to a genuine parametrized TQFT over $BG$, with a partition function independent of the connection.
More precisely, we first consider the partition function of the projective theory $Z_{\mathrm{proj}}(M^d,(g,A,W^{d+1}))$, obtained by projectivizing with respect to the universal scalar $\U(1)$.
To define it, let us consider a vector $Z_{\mathrm{proj}}(M^d,(g,A))$ given by the relative theory, which is an element of the one-dimensional vector space assigned to $(M^d,g,A)$ by the inflow invertible theory.
Then for a basis vector $e_W$ obtained from a bulk extension $W^{d+1}$ of $M^d$ together with an extension of the background $(g,A)$, we have $Z_{\mathrm{proj}}(M^d,(g,A))=Z_{\mathrm{proj}}(M^d,(g,A,W^{d+1}))\cdot e_W$.
In this setup, we can write the full partition function as
\begin{equation}
    Z(M^d,(g,A))=\exp\left(2\pi i\int_{W^{d+1}}\Omega(g,A,W^{d+1})\right)Z_{\mathrm{proj}}(M^d,(g,A,W^{d+1})),
\end{equation}
where $g$ is the classifying map and $A$ is the connection. Here we introduce a $(d+1)$-dimensional manifold $W^{d+1}$ whose boundary is the spacetime manifold $M^d$, and both $g$ and $A$ are also extended to $W^{d+1}$.
We write $\Omega\in\Omega_{\mathrm{cl}}^{d+1}(B_{\nabla}G)$ for the universal closed form and $\Omega(g,A,W^{d+1})=\widetilde g_{\nabla}^*\Omega$ for its pullback, where $\widetilde g_{\nabla}\colon W^{d+1}\to B_{\nabla}G$ extends $g_{\nabla}$.
When the corresponding TQFT is simple and its projective theory has only finitely many types of invertible topological operators, the projective moduli space $P\mathcal{M}$ is $\pi$-finite, and hence the projective partition function is independent of the connection.
In such a case, the dependence of the scalar factor of the partition function on the choice of the connection can be seen from the difference of the integrals of $\Omega(g,A,W^{d+1})$ over two different extensions $W^{d+1}_1$ and $W^{d+1}_2$ with the same boundary $M^d$:
\begin{equation}\label{eq:bulk-dependence_conti_qft}
    \int_{W^{d+1}_1}\Omega(g,A,W^{d+1}_1)-\int_{W^{d+1}_2}\Omega(g,A,W^{d+1}_2)=\langle g_{Y}^*\alpha_{\mathrm{QFT}},[W^{d+1}_1\cup_{M^d}\overline{W^{d+1}_2}]\rangle\in\mathbb{R}/\mathbb{Z},
\end{equation}
which is nonzero in general.
Here, $\alpha_{\mathrm{QFT}}\in\rH^{d+1}(BG,\mathbb{R}/\mathbb{Z})$ denotes the bulk response class determined by $\Omega$, and $g_Y$ is the classifying map on the glued manifold. 
This bulk dependence is also well-defined when the projective theory also depends on connections.
Since the original theory is non-anomalous and independent of the bulk extension, the bulk dependence of the projective partition function cancels that of the scalar factor involving $\Omega$.
When the $G$ symmetry is anomalous and has a torsion 't Hooft anomaly $g^*\omega$, the bulk dependence of the gauge invariant partition function $Z(M^d,(g,A))\exp(2\pi i\int_{W^{d+1}}g^*\omega)$ is the sum of the contributions from $\Omega$ and $g^*\omega$.

The corresponding picture from the perspective of parametrized TQFTs is as follows. Instead of considering a map $g\colon BG\rightarrow P\mathcal{M}$ to the projective moduli space, we consider a map
\begin{equation}
    B_{\nabla}G\xrightarrow{u} BG\rightarrow P\mathcal{M}.
\end{equation}
The $\U(1)$-valued anomaly $[g^*\omega]\in\rH^{d+1}(BG,\U(1)_{\mathrm{disc}})$ is an obstruction to lifting to $\mathcal{M}$.
Then the anomaly class is just given as $u^*[g^*\omega]\in \rH^{d+1}(B_{\nabla}G,\U(1)_{\mathrm{disc}})$. Since the pullback map $u^*\colon \rH^{d+1}(BG,\U(1)_{\mathrm{disc}})\rightarrow \rH^{d+1}(B_{\nabla}G,\U(1)_{\mathrm{disc}})$ is an isomorphism~\cite[Lemma~5.2]{Bunke:2016sheaves}, the anomaly classification is the same as that of the corresponding parametrized TQFT.

\bigskip
We now describe, in terms of differential cohomology, the relation between this bulk dependence \eqref{eq:bulk-dependence_conti_qft} and the $\U(1)$-valued obstruction class of the corresponding parametrized TQFT, which is an element of $\rH^{d+1}(B_{\nabla}G,\U(1)_{\mathrm{disc}})\cong \rH^{d+1}(BG,\U(1)_{\mathrm{disc}})$.
Given an anomaly class $[\alpha_{\mathrm{fam}}]\in\rH^{d+1}(BG,\U(1)_{\mathrm{disc}})$, we define a differential cohomology class~\cite{Cheeger:1985characters} $[\widehat{\alpha}_{\mathrm{fam}}]\in\widehat{\rH}^{d+2}(B_{\nabla}G,\mathbb{Z})$ by
\begin{equation}
    [\widehat{\alpha}_{\mathrm{fam}}]=j(u^*[\alpha_{\mathrm{fam}}])\in\widehat{\rH}^{d+2}(B_{\nabla}G,\mathbb{Z}),
\end{equation}
where $u\colon B_{\nabla}G\rightarrow BG$ is the forgetful map and $j\colon\rH^{d+1}(B_{\nabla}G,\U(1)_{\mathrm{disc}})\rightarrow\widehat{\rH}^{d+2}(B_{\nabla}G,\mathbb{Z})$ is the canonical inclusion of flat differential cohomology classes.
For a smooth manifold $X$, a class in $\widehat{\rH}^{d+2}(X,\mathbb{Z})$ can be represented by a triple $(N,h,F)\in C^{d+2}(X,\mathbb{Z})\times C^{d+1}(X,\mathbb{R})\times\Omega^{d+2}(X)$ such that
\begin{equation}
    \delta N=0,\quad \delta h=\iota(F)-N,\quad dF=0,
\end{equation}
where $\iota$ is the de Rham map.
Two such triples represent the same differential cohomology class if they are related by
\begin{equation}
    (N,h,F)\sim(N+\delta b,h-b-\delta c,F),
\end{equation}
where $b\in C^{d+1}(X,\mathbb{Z})$ and $c\in C^d(X,\mathbb{R})$.
In this description, when the anomaly in the standard topology vanishes, $[\widehat{\alpha}_{\mathrm{fam}}]$ is represented as
\begin{equation}
    [\widehat{\alpha}_{\mathrm{fam}}]=[(0,u^*\widetilde{\alpha}_{\mathrm{fam}},0)],
\end{equation}
where $\widetilde{\alpha}_{\mathrm{fam}}\in Z^{d+1}(BG,\mathbb{R})$ is a real lift of $\alpha_{\mathrm{fam}}$, identifying $\U(1)_{\mathrm{disc}}$ with $\mathbb{R}/\mathbb{Z}$.
On the other hand, the universal form $\Omega$ defines
\begin{equation}
    [\widehat{\alpha}_{\mathrm{QFT}}]=[(0,\iota(\Omega),0)]\in\widehat{\rH}^{d+2}(B_{\nabla}G,\mathbb{Z}).
\end{equation}
Let $\pi\colon\mathbb{R}\to\mathbb{R}/\mathbb{Z}$ be the quotient map and $\pi_*$ the induced map on cohomology.
By the definition of the equivalence relation, the two classes are the same if and only if
\begin{equation}
    \pi_*[\iota(\Omega)]
    =u^*[\alpha_{\mathrm{fam}}]
    \quad\text{in }\rH^{d+1}(B_{\nabla}G,\mathbb{R}/\mathbb{Z}).
\end{equation}
To see this explicitly, let us assume that the anomaly class is given by $[\alpha_{\mathrm{fam}}]=\pi_*[\lambda p]$ and the scalar response is given by $\Omega=\lambda\omega_p$.
Here $[p]\in\rH^{d+1}(BG,\mathbb{Z})$ is an integral characteristic class, $\lambda\in\mathbb{R}$, and $\omega_p\in\Omega_{\mathrm{cl}}^{d+1}(B_{\nabla}G)$ is the universal Chern--Weil form representing the real image of $u^*[p]$.
Let $p\in Z^{d+1}(BG,\mathbb{Z})$ be a cocycle representative of this class.
Then the Chern--Weil formula~\cite[Thm.~2.2]{Cheeger:1985characters} implies that there is a real cochain $h_p\in C^d(B_{\nabla}G,\mathbb{R})$ such that
\begin{equation}
    \delta h_p=\iota(\omega_p)-u^*p.
\end{equation}
Taking $\widetilde{\alpha}_{\mathrm{fam}}=\lambda p$, we obtain
\begin{equation}
    \iota(\Omega)-u^*\widetilde{\alpha}_{\mathrm{fam}}
    =\delta(\lambda h_p),
\end{equation}
and so
\begin{equation}
    [\widehat{\alpha}_{\mathrm{QFT}}]=[\widehat{\alpha}_{\mathrm{fam}}]\in\widehat{\rH}^{d+2}(B_{\nabla}G,\mathbb{Z}).
\end{equation}
Thus, in this case, the two descriptions give the same differential cohomology class.

To see the bulk dependence, consider a closed oriented manifold $Y^{d+1}$ with background $g_{Y,\nabla}\colon Y^{d+1}\to B_{\nabla}G$, and let $g_Y=u\circ g_{Y,\nabla}$.
Then we have
\begin{equation}\label{eq:bulk-dependence_conti_family}
    \exp\left(2\pi i\int_{Y^{d+1}}g_{Y,\nabla}^*\Omega\right)
    =\exp\left(2\pi i\left\langle g_Y^*\alpha_{\mathrm{fam}},[Y^{d+1}]\right\rangle\right).
\end{equation}
Here $Y^{d+1}=W^{d+1}_1\cup_{M^d}\overline{W^{d+1}_2}$ is the gluing of two extensions.
Comparing \eqref{eq:bulk-dependence_conti_qft} and \eqref{eq:bulk-dependence_conti_family}, we see that the bulk dependence of the scalar factor of the partition function is the same as the $\U(1)$-valued obstruction class of the corresponding parametrized TQFT with the discrete topology on $\U(1)$.

When the $G$ symmetry also has a torsion 't Hooft anomaly $\omega\in Z^{d+1}(BG,\mathbb{R}/\mathbb{Z})$, the corresponding differential cohomology class is given by \begin{equation}
    j(u^*[\omega])=[(u^*N_{\omega},u^*\widetilde{\omega},0)]\in\widehat{\rH}^{d+2}(B_{\nabla}G,\mathbb{Z}),
\end{equation}
where $\widetilde{\omega}\in C^{d+1}(BG,\mathbb{R})$ is the lift and $N_{\omega}\coloneqq -\delta\widetilde{\omega}\in Z^{d+2}(BG,\mathbb{Z})$.
Combining this with the contribution from $\alpha_{\mathrm{fam}}$, we can represent the total anomaly class by
\begin{equation}
    [(u^*N_{\omega},u^*\widetilde{\omega}+u^*\widetilde{\alpha}_{\mathrm{fam}},0)]\in\widehat{\rH}^{d+2}(B_{\nabla}G,\mathbb{Z}).
\end{equation}

\subsubsection{\texorpdfstring{$\U(1)$}{U1} symmetry in quantum mechanics}
As a simple example, let us consider a $(0+1)$D $N$-fold degenerate quantum mechanical system with a $\U(1)$ symmetry. Let $\rho\colon\U(1)\to\U(N)$ be its representation on the ground states. 
Write a background $\U(1)$ connection as $A$ and the induced $\U(N)$ connection $a$ as $\rho_*(A)$. Specifically, if the $\U(1)$ representation on the ground states is given by $\U(1)$ charges $q_i\in\mathbb Z$ for $i=1,\ldots,N$, then the induced $\U(N)$ connection is given by
\begin{equation}
    a\coloneqq \mathrm{diag}(q_1A,\ldots,q_N A).
\end{equation}
The partition function on the time circle is the trace of the holonomy of the background $\U(1)$ connection $A$ in this representation:
\begin{equation}\label{eq:qm-holonomy}
    Z(S^1,A)
    =\operatorname{tr}\left(\operatorname{Hol}_a(S^1)\right)
    =\sum_{i=1}^{N}\exp\left(iq_i\int_{S^1}A\right).
\end{equation}
Here $\operatorname{Hol}_a(S^1)=\exp(i\int_{S^1}a)$ denotes the holonomy of $a$. 
The connection $A$ is obtained by pulling back the universal connection on the universal $\U(1)$ bundle over $B_{\nabla}\U(1)$ along a map $M^1\rightarrow B_{\nabla}\U(1)$. We note that now the bundle class is trivial because $M^1=S^1$ is one-dimensional.
This partition function is well-defined and invariant under gauge transformations of the connection: $A\mapsto A+d\lambda$, for any $\mathbb{R}/2\pi\mathbb{Z}$-valued function $\lambda$.

Now we can say that the background gauge field $A$ is coupled to the system through the induced $\U(N)$ connection $a$.
On the other hand, the actual physical symmetry is given only by the projective representation $\tilde{\rho}\colon \U(1)\to\PU(N)$, so we decompose the connection as
\begin{equation}
    a=a_0+\frac{\operatorname{tr}(a)}{N}I,\quad \operatorname{tr}(a_0)=0.
\end{equation}
Explicitly, we have
\begin{equation}
    a_0=\mathrm{diag}\left(q_i-q\right)A,\quad q=\frac{1}{N}\sum_{i=1}^{N}q_i.
\end{equation}
We call this $q$ the average charge of the $\U(1)$ symmetry.
Then the partition function is written as
\begin{equation}
    Z(S^1,A)=\operatorname{tr}\operatorname{Hol}_{a_0}(S^1)\exp\left(iq\int_{S^1}A\right).
\end{equation}
The last factor is generally not gauge invariant and can also change under the shift $a\mapsto a+bI$, where $b$ is a closed $1$-form. This shift leaves the $\PU(N)$ bundle class unchanged.
Such an ambiguity is controlled by the bulk phase factor
\begin{equation}
    \exp\left(2\pi iq\int_{B^2}\frac{F_A}{2\pi}\right),
\end{equation}
where $B^2$ is a two-dimensional manifold with boundary $S^1$, the time circle of the quantum mechanical system, and $F_A$ is the field strength of the background $\U(1)$ gauge field.
The relative Chern--Weil formula~\cite[Sec.~1 and Thm.~2.2]{Cheeger:1985characters} reads
\begin{equation}
    \int_{B^2}\frac{F_A}{2\pi}=\langle c_1(L,\tau),[B^2,S^1]\rangle+\int_{S^1}\frac{A}{2\pi},
\end{equation}
where $L\rightarrow B^2$ is the extension of the $\U(1)$ bundle, and $\tau$ is a trivialization of $L$ over $S^1$. The first term is integer-valued.
Take two different extensions $B^2_1$ and $B^2_2$ of the same $\U(1)$ bundle.
The ratio of the corresponding bulk phase factors is given by
\begin{equation}
    \exp\left(2\pi iq\int_{Y^2}\frac{F_A}{2\pi}\right)=\exp\left(2\pi iq\langle c_1(L_{Y^2}),[Y^2]\rangle\right),\quad Y^2=B^2_1\cup_{S^1} \overline{B^2_2}.
\end{equation}
Since the average charge $q$ need not be an integer, this bulk phase can be nontrivial. The corresponding anomaly is $qc_1\in\rH^2(B\U(1),\U(1)_{\mathrm{disc}})$.

\paragraph{Family picture}
Let us describe the above quantum mechanical example from the perspective of parametrized TQFTs.
The moduli space of an $N$-fold degenerate $(0+1)$D quantum mechanical system is given by $B\U(N)$, and thus the anomaly is computed by studying the following diagram:
\begin{equation}
    \begin{tikzcd}
        B\U(1)
            \arrow[r]
        &
        B\U(N)
            \arrow[d]
        \\
        X
            \arrow[r]
        &
        B\PU(N)\mathrlap{.}
    \end{tikzcd}
\end{equation}
Let us consider a $\U(1)$ symmetry acting on the system. Then the parameter space $X$ is given by $B_{\nabla}\U(1)$.
If we take the standard topology on $\U(1)$, we always have a lift of the map $X\rightarrow B\PU(N)$ to $B\U(N)$ because $\rH^3(X,\mathbb{Z})\cong\rH^3(B\U(1),\mathbb{Z})\cong\rH^3(K(\mathbb{Z},2),\mathbb{Z})=0$.
This agrees with the usual classification of 't Hooft anomalies, since we usually say that there is no 't Hooft anomaly for a $\U(1)$ symmetry in $(0+1)$D quantum mechanics. This is because the 't Hooft anomaly of a $G$ symmetry is classified by projective representations of $G$, and every projective representation of $\U(1)$ lifts to a linear representation.
This classification is also understood by the fact that the space of invertible theories in $(1+1)$D with a $\U(1)$ symmetry is connected. Indeed, all their partition functions take the form $\exp\left(i\theta\int\frac{F}{2\pi}\right)$, where $F$ is the field strength of the background $\U(1)$ gauge field, and any $\theta$ can be continuously deformed to $0$.

Let us next consider lifting the map $X\to B\PU(N)$ to $B\SU(N)$. Since the fiber of $B\SU(N)\to B\PU(N)$ is $B\mathbb Z_N$, the obstruction takes values in $\rH^2(X,\mathbb Z_N)$.
The relation to the anomaly class in $\rH^2(X,\U(1)_{\mathrm{disc}})$ will become clear shortly.
To see the physical meaning of this obstruction class, let us consider a $\U(1)$ symmetry acting on an $N$-dimensional ground-state Hilbert space.
Then the action of the $\U(1)$ symmetry on the ground-state Hilbert space defines a map
\begin{equation}
    \tilde{\rho}\colon \U(1)\rightarrow \PU(N)=\U(N)/\U(1).
\end{equation}
Since $\rH_{\mathrm{gp}}^2(\U(1),\U(1))= 0$, we always have a lift to $\U(N)$.
Now the anomaly on this $(0+1)$D system is given by the average $\U(1)$ charge of the ground states: $q=\sum_{i}q_i/N$, where $q_i\in\mathbb{Z}$ are the $\U(1)$ charges of the ground states.
To be precise, this average charge detects an obstruction to the existence of a lift to $\SU(N)$, not $\U(N)$. Note that $\SU(N)$ fits into the following exact sequence
\begin{equation}
    1\rightarrow \mathbb{Z}_N \rightarrow \SU(N) \rightarrow \PU(N) \rightarrow 1.
\end{equation}
When $q=k/N$, the obstruction class is given by
\begin{equation}
    kc_1 \in \rH^2(B\U(1),\mathbb{Z}_N)\cong \mathbb{Z}_N.
\end{equation}
Again, we would not call this an 't Hooft anomaly of the $\U(1)$ symmetry: the obstruction to lifting to $\U(N)$ with the standard topology vanishes, since $\rH^3(B\U(1),\mathbb{Z})=0$.
From the field-theoretic perspective described above, this obstruction is represented by the bulk symmetry-protected topological (SPT) action $(k/N)c_1$. Mathematically, this correspondence gives
\begin{equation}
    \rH^2(B\U(1),\mathbb{Z}_N)\hookrightarrow \rH^2(B\U(1),\varinjlim_M\mathbb{Z}_M)\cong \rH^2(B\U(1),\mathbb{Q}/\mathbb{Z});\quad kc_1\mapsto \frac{k}{N}c_1.
\end{equation}
The image under this map naturally defines an element of $\rH^2(B\U(1),\U(1)_{\mathrm{disc}})$ through the inclusion $\varinjlim_M\mathbb{Z}_M\cong \mathbb{Q}/\mathbb{Z}\hookrightarrow\U(1)_{\mathrm{disc}}$.
The resulting class is the obstruction to making the local $\SU(N)$ lifts satisfy the cocycle condition by multiplying them by locally constant $\U(1)$ phases.

\subsubsection{Fractional quantum Hall response}
Another similar example is found in $(2+1)$D TQFTs with a nontrivial fractional quantum Hall response. For instance, consider the $\U(1)_{k}$ Chern--Simons theory with $k>0$ coupled to a background $\U(1)$ gauge field $A$ as
\begin{equation}
    2\pi S=\frac{k}{4\pi}\int a\wedge da + \frac{1}{2\pi}\int A\wedge da.
\end{equation}
Here, $a$ is a dynamical $\U(1)$ gauge field, and $A$ is the background $\U(1)$ gauge field.
To define the theory on oriented manifolds, the level $k$ must be an even integer. Note that this theory also has a framing anomaly because the chiral central charge is $c_-=1$. 

Let $L\to M^3$ be the background $\U(1)$ bundle with connection $A$. In the following discussion, we keep the framing fixed throughout.
The full partition function $Z(M^3,A)$ is well-defined and invariant under gauge transformations of $A$.
To see its dependence on the connection and on the underlying bundle, consider a shift $A\mapsto A+b$ by a global $1$-form $b$.
The change of integration variable $a'=a+b/k$ is allowed within each dynamical bundle and we obtain
\begin{equation}\label{eq:fqh-connection-response}
    Z(M^3,A+b)=
    \exp\left(-\frac{i}{4\pi k}\int_{M^3}
    \left(2b\wedge F_A+b\wedge db\right)\right)Z(M^3,A).
\end{equation}
This is precisely the variation of the fractional Chern--Simons response, whose local expression is $\exp(-i\int_{M^3}A\wedge dA/(4\pi k))$.
However, this local expression does not by itself define a complex number for an arbitrary background bundle.

To define the fractional Chern--Simons term unambiguously, choose an extension of $(L,A)$ to an oriented four-manifold $W^4$ with $\partial W^4=M^3$.%
\footnote{Such an extension always exists because $\Omega_{3}^{\SO}(B\U(1))=0$.}
Then the partition function is written as
\begin{equation}\label{eq:fqh-response-separation}
    Z(M^3,A)=Z_{\mathrm{top}}(M^3,(L,W^4))
    \exp\left(-2\pi i\,\frac{1}{2k}
    \int_{W^4}\left(\frac{F_A}{2\pi}\right)^2\right).
\end{equation}
Now the term $Z_{\mathrm{top}}$ is independent of the connection on the fixed bundle $L$, whereas in the QM example the remaining factor generally still depends on $A$.
The bulk exponential has exactly the variation in \eqref{eq:fqh-connection-response}, as follows by extending $b$ to $W^4$ and applying Stokes' theorem.
We can then get rid of the connection dependence and make the partition function topological by adding a local fractional Chern--Simons term.
However, the remaining topological factor $Z_{\mathrm{top}}$ depends on the choice of extension $W^4$.
Specifically, take two extensions $(W^4_1,L_1,A_1)$ and $(W^4_2,L_2,A_2)$ of the same boundary bundle with connection $(L,A)$.
Let $Y^4=W^4_1\cup_{M^3}\overline{W^4_2}$ and let $L_Y$ be the glued bundle.
Now the relative Chern--Weil formula~\cite[Sec.~1 and Thm.~2.2]{Cheeger:1985characters} reads
\begin{equation}\label{eq:fqh-relative-chern-weil}
    \int_{W^4}\left(\frac{F_A}{2\pi}\right)^2
    =\left\langle c_1^2(L,\tau),[W^4,M^3]\right\rangle
    +\mathrm{CS}_{c_1^2}(M^3,A,\tau),
\end{equation}
where $\tau$ is a trivialization of $c_1(L)^2$ on $M^3$, and $c_1^2(L,\tau)\in\rH^4(W^4,M^3;\mathbb Z)$ is the resulting relative class.
Such a trivialization exists because $M^3$ is three-dimensional. When $L|_{M^3}$ is trivial and $\tau$ is induced by a trivialization of this bundle, the second term on the right-hand side is written as
\begin{equation}\label{eq:fqh-local-chern-simons}
    \mathrm{CS}_{c_1^2}(M^3,A,\tau)
    =\frac{1}{4\pi^2}\int_{M^3}A\wedge dA.
\end{equation}
Thus, the ratio of the two scalar bulk factors with different extensions is given by
\begin{equation}\label{eq:fqh-bulk-comparison}
    \exp\left(-2\pi i\,\frac{1}{2k}
    \int_{Y^4}\left(\frac{F_A}{2\pi}\right)^2\right)
    =\exp\left(-2\pi i\,\frac{1}{2k}
    \left\langle c_1(L_Y)^2,[Y^4]\right\rangle\right).
\end{equation}
Since the Chern number on $Y^4$ is an integer, this phase is nontrivial in general.
The scalar bulk factor represents $-c_1^2/(2k)\in\rH^4(B\U(1),\U(1)_{\mathrm{disc}})$.
The remaining factor $Z_{\mathrm{top}}$ acquires the inverse phase, so the full partition function is independent of the choice of extension.

Let us briefly comment on the fractional Chern--Simons term.
The term is derived using local response theory, and its local expression assumes a trivialization of the background bundle.
However, a necessary condition for the partition function to be nonzero is given by
\begin{equation}
    \frac{1}{k}\int\frac{F_A}{2\pi}\in\mathbb{Z}.
\end{equation}
This is consistent with the fact that the partition function would vanish if the condition is not satisfied, since in that case the $\U(1)$ gauge field traps a nontrivial anyon line operator, whose $1$-form symmetry is spontaneously broken.
Even under this constraint, one can still have a nontrivial gauge field on an oriented $4$-manifold $W^4\cup_{M^3}\overline{W^{\prime 4}}$ such that the integral of $c_1^2(A)$ over it is one.

\subsubsection{Framing anomaly revisited}\label{sec:framing_anomaly}
Now let us discuss again the framing anomaly of $(2+1)$D TQFTs as in the original work of Witten~\cite{Witten:1988hf}.
Recall that the framing anomaly of a $(2+1)$D TQFT is an obstruction to making the theory simultaneously independent of the metric and the framing.
Following the above formulation, we use $B_{\nabla}\SO(3)$ to describe the background given by the oriented orthonormal frame bundle of $TM^3$ with the Levi--Civita connection of the metric $g$.
More specifically, the partition function of Chern--Simons theory defined naively by the path integral of the gauge field actually depends on the metric $g$, yet the combination 
\begin{equation}
    Z_{\mathrm{CS}}(M^3,g)\exp\left(-\frac{2\pi i}{24}c_-\int_{M^3}\mathrm{CS}_{\mathrm{grav}}\right)
\end{equation}
is independent of the metric, where $\mathrm{CS}_{\mathrm{grav}}$ is the gravitational Chern--Simons term and $c_-$ is the chiral central charge of the TQFT.
Since this is a local counterterm, we would not say that the system has a gravitational anomaly.

In our formalism, the partition function of the Chern--Simons theory is decomposed as
\begin{equation}
    Z_{\mathrm{CS}}(M^3,g)=\exp\left(\frac{2\pi i}{24}c_-\int_{W^4}p_1(\nabla_W)\right)Z_{\mathrm{proj}}(M^3,W^4).
\end{equation}
Here we choose a trivialization $\tau$ of $p_1(TM^3)$~\cite[Appendix~B]{Blanchet:1995tqft}, which exists because $M^3$ is three-dimensional.
Near the boundary, we take a neighborhood $M^3\times[0,\epsilon)$ with the product metric $g+dt^2$, where $t$ is the normal coordinate.
We choose $\nabla_W$ on $TW^4$ to agree there with the Levi--Civita connection of this product metric, i.e., the Levi--Civita connection of $g$ along $M^3$ and the trivial connection in the normal direction.
The first Pontryagin form is $p_1(\nabla_W)=-\frac{1}{8\pi^2}\operatorname{tr}(R_W\wedge R_W)$, where $R_W$ is the curvature of $\nabla_W$ and the trace is taken in the vector representation~\cite{Chern:1974characteristic}.
On the right-hand side, the metric dependence only appears in the first factor.
To see the bulk dependence of the first factor, we use the relative Chern--Weil formula~\cite[Sec.~1 and Thm.~2.2]{Cheeger:1985characters}, which reads
\begin{equation}\label{eq:grav-relative-chern-weil}
    \int_{W^4}p_1(\nabla_W)
    =\left\langle p_1(TW^4,\tau),[W^4,M^3]\right\rangle
    +\int_{M^3}\mathrm{CS}_{\mathrm{grav}},
\end{equation}
where $p_1(TW^4,\tau)\in\rH^4(W^4,M^3;\mathbb Z)$ is the relative Pontryagin class, and the last term is the real lift of the gravitational Chern--Simons invariant determined by $g$ and $\tau$.
For fixed $g$ and $\tau$, this last term is independent of the choice of extension $W^4$.
However, there is no canonical choice of such a trivialization compatible with cutting and gluing, and so the partition function of the TQFT depends on the choice of the trivialization.
If the counterterm is instead defined using the bulk integral alone, the ratio of the two bulk exponential factors is given by
\begin{equation}
    \exp\left(-\frac{2\pi i}{24}c_-\left\langle p_1(TY^4),[Y^4]\right\rangle\right),
    \qquad Y^4=W^4_1\cup_{M^3}\overline{W^{4}_2},
\end{equation}
which is nontrivial in general.

\subsection{Dynamical constraints}\label{sec:dynamical_constraint}
The anomaly we have discussed in previous sections is based on the parametrized TQFT picture, or TQFTs coupled to parameter spaces, where we already specify the IR theory for global symmetries.
In this sense, the anomaly here is determined not only by kinematic properties at the UV scale, but also by the dynamical properties at the IR scale.
For this reason, we would distinguish the anomaly here from the usual 't Hooft anomaly, and call it \textit{an anomaly on gapped theories}.
This picture is somewhat different from the usual anomaly matching, where we only consider the kinematic properties at the UV scale, and the IR theory is not specified.
As a consequence, this anomaly does not by itself constrain which IR theory is realized, since the IR theory has already been specified.
Indeed, the fractional quantum Hall effect is realized by an on-site $\U(1)$ global symmetry, which is non-anomalous.

What we can say from the anomaly on gapped theories is rather trivial. For instance, let us consider again the baby example of an $N$-fold degenerate $(0+1)$D quantum mechanical system with a $\U(1)$ global symmetry. 
The anomaly on the system is given by the average $\U(1)$ charge of the ground states.
Instead of fixing the dimension of the ground-state Hilbert space, let us fix the average $\U(1)$ charge of the ground states to be a rational number $\nu=n/m$.
Then we can say that the dimension of the ground-state Hilbert space must be a multiple of $m$, if the system is gapped.

This dynamical constraint is rather trivial and not so useful. To obtain a more nontrivial dynamical constraint, let us consider a $(1+1)$D system with  global $\U(1)$ and translation symmetries.
Such a system can have a Lieb--Schultz--Mattis (LSM) type constraint~\cite{Lieb:1961fr,Oshikawa:2000zwq,Hastings:2003zx}, which states that if the filling of the $\U(1)$ charge per unit cell is a noninteger rational number $\nu=n/m$, then the ground state of the corresponding sector must be either gapless or degenerate.
In particular, if the system is gapped, then the ground-state degeneracy must be a multiple of $m$.

We first note that there are several types of LSM-type constraints in $(1+1)$D. A typical example arises when the Hilbert space of each unit cell carries a nontrivial projective representation of an internal symmetry $G$. With a translation symmetry, such a system also carries an LSM-type constraint, which states that the system cannot have a unique gapped ground state.
An important difference between this LSM-type constraint and the one discussed above is that the former forbids any unique gapped ground state within the total tensor-product Hilbert space, while the latter only forbids a unique gapped ground state within a fixed sector with a fractional filling.
Then a crucial observation is that this original LSM constraint is \textit{not} understood as an anomaly between the microscopic $\U(1)$ and translation symmetries in the usual sense~\cite{Else:2020jln,Cheng:2022sgb}; the formal calculation of the anomaly classification only involves the perturbative anomaly: $\rH^4(B\U(1)\times B\mathbb{Z},\mathbb{Z})\cong \rH^4(B\U(1),\mathbb{Z})\cong \mathbb{Z}$. Here, we take the standard topology on $\U(1)$ in the calculation, i.e., we consider the deformation equivalence of anomalies.
This is not a contradiction because this cohomology classification only describes which anomaly we have in the whole Hilbert space, and does not say anything about the anomaly in a fixed sector with a fractional filling.

How, then, should we understand anomalies in a fixed sector, and can the original LSM constraint be understood in terms of anomaly matching?
As discussed in \cite{Cheng:2022sgb}, the answer is yes. In our framework, this is understood by taking the discrete topology on $\U(1)$, i.e., considering the isomorphism class of anomalies.
The corresponding bulk SPT response would be
\begin{equation}
    \exp\left(2\pi i\nu\int C\cup c_1\right),
\end{equation}
where $C$ is a background gauge field of the translation symmetry, and $c_1$ is the first Chern class of the background $\U(1)$ gauge field.
The point is that once one fixes the filling $\nu$, one cannot continuously deform the bulk SPT theta-like term anymore, and so we need to consider the isomorphism class of the anomaly, not the deformation class. 
The bulk SPT term is now a nontrivial element of $\rH^3(B\U(1)\times B\mathbb{Z},\mathbb{Q}/\mathbb{Z})\cong\mathbb{Q}/\mathbb{Z}$, from which we have a natural inclusion into $\rH^3(B\U(1)\times B\mathbb{Z},\U(1)_{\mathrm{disc}})$.

\section{Anomaly constraints on ground-state degeneracies}\label{sec:anomaly_constraint}
In this section, we discuss some applications of anomalies on TQFTs coupled to parameter spaces.
We first provide a general formula relating the anomaly to the ground-state degeneracy of a gapped system. Then we give some applications of the formula for $(2+1)$D TQFTs and symmetry-preserving TQFTs. We also discuss how general anomalies can be detected by $(d+1)$-dimensional manifolds whose fibers are closed $(d-1)$-dimensional spatial manifolds.

\subsection{Anomaly-GSD formula}\label{sec:anomaly-gsd}
According to the Atiyah--Segal axioms of (T)QFTs~\cite{Atiyah:1989vu,Segal:2002ei}, we assign a vector space to every closed spatial $(d-1)$D manifold $\Sigma^{d-1}$. Let us denote the vector space by $Z(\Sigma^{d-1})$.
Take a based loop $\gamma$ in $B\operatorname{Diff}(\Sigma^{d-1})$, and let $f\colon\Sigma^{d-1}\to\Sigma^{d-1}$ be a gluing diffeomorphism for the associated $\Sigma^{d-1}$-bundle over $S^1$. Then we have a partition function
\begin{equation}
    Z(\Sigma^{d-1}\times_{f}S^1),
\end{equation}
which is more explicitly given as 
\begin{equation}
    \operatorname{tr}_{Z(\Sigma^{d-1})}\left(U_\gamma\right),
\end{equation}
where $U_\gamma$ is the unitary operator obtained by parallel transport in the ground-state bundle around the loop $\gamma$.
In particular, when considering a contractible loop $\gamma$ in $B\operatorname{Diff}(\Sigma^{d-1})$, namely taking a trivial element in $\pi_1(B\operatorname{Diff}(\Sigma^{d-1}))\cong\pi_0(\operatorname{Diff}(\Sigma^{d-1}))$, we have a partition function on $\Sigma\times S^1$.

Now let us consider the same theory with a map to our parameter space $X$. For a contractible loop $\gamma$, one can similarly have 
\begin{equation}
    Z(\Sigma^{d-1}\times S^1,g\colon \Sigma^{d-1}\times S^1\to X).
\end{equation}
We know that the phase ambiguity of the partition function on $\Sigma^{d-1}\times S^1$ is given by anomaly inflow from an invertible TQFT in $(d+1)$D, whose partition function is written as $\exp(2\pi i\int g^*\omega)$. 
Namely, the combination
\begin{equation}
    Z(\Sigma^{d-1}\times S^1,g\colon \Sigma^{d-1}\times S^1\to X)\,\exp\left(2\pi i\int_{W^{d+1}}g^*\omega\right)
\end{equation}
has no phase ambiguity. 
Here, $W^{d+1}$ is a fibered space
\begin{equation}
    \Sigma^{d-1}\rightarrow W^{d+1} \rightarrow Y^2
\end{equation} 
such that $\partial W^{d+1}=\Sigma^{d-1}\times S^1$. Here, we consider anomalies when taking the discrete topology on $\U(1)$, not the standard topology. See Sec.~\ref{sec:anomaly_parameter} for the difference between the two choices of topology on $\U(1)$.
When we consider a theory coupled to a background gauge field for a continuous $G$ symmetry, the parameter space $X$ is given by $B_{\nabla}G$.
In such a case, we further write the above combination as
\begin{equation}
    Z_{\mathrm{proj}}(\Sigma^{d-1}\times S^1,g_{\nabla}\colon \Sigma^{d-1}\times S^1\to B_{\nabla}G)\,\exp\left(2\pi i\int_{W^{d+1}}g^*\omega
    +2\pi i\int_{W^{d+1}}\Omega(g,A,W^{d+1})\right),
\end{equation}
as discussed in Sec.~\ref{sec:conti_sym}. 
In the following, we use $\omega$ to denote the total anomaly, including both contributions.

Let us consider the $(0+1)$D theory obtained by compactifying the original theory on $\Sigma^{d-1}$ and denote its partition function by $Z_{\Sigma^{d-1}}(S^1,g^\flat)$. Here $g^\flat$ is an element of $\operatorname{Map}(S^1,\operatorname{Map}(\Sigma^{d-1},X))$ defined by the adjunction $\operatorname{Map}(\Sigma^{d-1}\times S^1,X)\cong\operatorname{Map}(S^1,\operatorname{Map}(\Sigma^{d-1},X))$. Since the compactified theory is a $(0+1)$D theory, we now have a projective unitary bundle over $S^1$, whose rank is the ground-state degeneracy $N(\Sigma^{d-1},\hat{g})$ of the original theory, where $\hat{g}\coloneqq g^\flat(\ast)$ for a chosen point $\ast\in S^1$.
An anomaly of this theory is canceled by an SPT action $\Omega_{\Sigma}$ on the $(1+1)$D bulk $Y^2$, i.e., the combination $Z_{\Sigma^{d-1}}(S^1,g^\flat)\exp(2\pi i\int\Omega_{\Sigma})$ has no phase ambiguity.

The fractional phase of this projective bundle is $N(\Sigma^{d-1},\hat{g})$-torsion~\cite[Prop.~2.1(iv)]{Atiyah:2004jv}%
\footnote{We use unitary notation for convenience, but the argument does not require unitarity. For a general complex-linear TQFT, one replaces $\U(N)$, $\PU(N)$, and $\SU(N)$ by $\mathrm{GL}_N(\mathbb{C})$, $\mathrm{PGL}_N(\mathbb{C})$, and $\mathrm{SL}_N(\mathbb{C})$, respectively. The kernel of $\mathrm{SL}_N(\mathbb{C})\to\mathrm{PGL}_N(\mathbb{C})$ is again $\mathbb{Z}_N$, so the same $N$-torsion argument applies.}, as explained below, and hence $\Omega_{\Sigma}$ satisfies
\begin{equation}
    N(\Sigma^{d-1},\hat{g})\int_{Y^2}\Omega_{\Sigma}\in\mathbb{Z}
\end{equation}
for any oriented closed two-dimensional manifold $Y^2$. 
On the other hand, anomaly matching gives
\begin{equation}
    \Omega_{\Sigma}=\int_{\Sigma^{d-1}}g^*\omega.
\end{equation}
Hence, we obtain the following constraint.
\begin{prop}\label{prop:anomaly-gsd}
Consider a $d$-dimensional TQFT coupled to a parameter space $X$, and let $\omega$ be its $\U(1)$-valued bosonic torsion anomaly, with the universal $\U(1)$ given the discrete topology.
Let $\Sigma^{d-1}\rightarrow W^{d+1}\rightarrow Y^2$ be a bundle with closed fiber $\Sigma^{d-1}$ over a closed connected surface $Y^2$, with oriented total space $W^{d+1}$, and let $g\colon W^{d+1}\rightarrow X$ be the parameter map. The anomaly gives the bulk phase $\exp(2\pi i\int_{W^{d+1}}g^*\omega)$.
Write $\hat{g}=g|_{\Sigma^{d-1}}$ for the restriction to a fiber, and let $N(\Sigma^{d-1},\hat{g})$ denote the ground-state degeneracy on that fiber with background $\hat{g}$. Then
\begin{equation}\label{eq:anomaly-gsd}
    N(\Sigma^{d-1},\hat{g})\int_{W^{d+1}}g^*\omega\in\mathbb{Z}.
\end{equation}
\end{prop}

Let us comment on the $(1+1)$D SPT $\Omega_{\Sigma}$ for the compactified theory. Let $P$ be its $\PU(N)$-bundle of projective ground states over the parameter space $\mathcal{P}=\operatorname{Map}(\Sigma^{d-1},X)\mathord{\sslash} \operatorname{Diff}(\Sigma^{d-1})$.%
\footnote{For $X=B_{\nabla}G$, $\operatorname{Map}$ denotes the mapping stack, and the quotient is understood as a homotopy quotient.}
We may describe the SPT action by a differential cohomology class $\widehat{\Omega}_{\Sigma}\in\widehat{\rH}^3(\mathcal{P},\mathbb{Z})$, whose surface holonomy gives the bulk phase. Applying the characteristic class map $I$, we obtain
\begin{equation}
    \operatorname{DD}(P)=-I(\widehat{\Omega}_{\Sigma})\in \rH^3(\mathcal{P},\mathbb{Z}).
\end{equation}
This is known as the Dixmier--Douady class of the projective bundle, and its order divides $N$.
To see this, consider the obstruction $u(P)\in \rH^2(\mathcal{P},\mathbb{Z}_N)$ to lifting $P$ to an $\SU(N)$-bundle. Then $\operatorname{DD}(P)=\beta_N u(P)$, where $\beta_N$ is the Bockstein map for $0\to\mathbb{Z}\xrightarrow{\times N}\mathbb{Z}\to\mathbb{Z}_N\to0$.
The local $\SU(N)$ lifts define the phase cocycle $\Omega_P=u(P)/N\in \rH^2(\mathcal{P},\mathbb{R}/\mathbb{Z})$. Since $u(P)$ is $N$-torsion, both $\operatorname{DD}(P)$ and $\Omega_P$ are $N$-torsion. The physical assumption is that the bulk phase $\exp(2\pi i\int_{Y^2}\Omega_{\Sigma})$ equals the surface holonomy of the flat differential cohomology class represented by $\Omega_P$. This holonomy is an $N$-th root of unity, and so $N\int_{Y^2}\Omega_{\Sigma}\in\mathbb{Z}$.
When $Y^2$ is nonorientable, the same $N$-torsion argument applies with local coefficients twisted by $w_1(Y^2)$~\cite[Remark~6.2]{Hekmati:2016ikh}.

\subsection{Applications to 3D TQFTs}\label{sec:application_3d_tqft}
Consider a $(2+1)$-dimensional TQFT specified by a modular tensor category $\mathcal{B}$, and take a parameter space to be $B^2\mathbb{Z}_n\simeq K(\mathbb{Z}_n,2)$.
The anomaly on this TQFT is given by the following bulk SPT action~\cite{Hsin:2018vcg}
\begin{equation}
    \nu\int_{W^4}\mathcal{P}(B) + \frac{c_-}{24}\int_{W^4}p_1(TW^4).
\end{equation}
Here, $B$ is the background gauge field for the $\mathbb{Z}_n$ $1$-form symmetry, and $\mathcal{P}$ denotes the Pontryagin square operation.
For even $n$, the Pontryagin square is defined as
\begin{equation}
    \mathcal{P}\colon \rH^2(X,\mathbb{Z}_n)\rightarrow \rH^4(X,\mathbb{Z}_{2n}),\quad \mathcal{P}(B)=\tilde{B}\cup \tilde{B}+\tilde{B}\cup_1 \delta\tilde{B} \mod 2n,
\end{equation}
where $\tilde{B}$ is an integral lift of $B$. For odd $n$, we set $\mathcal{P}(B)=B\cup B\in\rH^4(X,\mathbb{Z}_n)$. In both cases, it satisfies
\begin{equation}
    \mathcal{P}(B_1+B_2)=\mathcal{P}(B_1)+\mathcal{P}(B_2)+2B_1\cup B_2.
\end{equation}
Then an anomaly of the $\mathbb{Z}_n$ $1$-form symmetry is classified by
\begin{equation}
    \nu=\frac{k}{\gcd(n,2)n},\quad k\in \mathbb{Z}_{\gcd(n,2)n}.
\end{equation}
Applying the general result of the previous section, we have a constraint on the ground-state degeneracy $N(\Sigma^2,\hat{B})$ of the TQFT on a closed oriented surface $\Sigma^2$ with a flux insertion $\hat{B}=B|_{\Sigma^2}$ of abelian anyons:
\begin{equation}
    N(\Sigma^2,\hat{B})\left(\nu\int_{W^4}\mathcal{P}(B) + \frac{c_-}{24}\int_{W^4}p_1(TW^4)\right) \in \mathbb{Z}.
\end{equation}
Let us write the spatial manifold as $\Sigma_g^2$ for a closed oriented surface of genus $g$.
Then for any flux insertion $\hat{B}$ with $g\geq 3$, one can find a configuration of $W^4$ such that $\int_{W^4}\mathcal{P}(B)=0$.%
\footnote{\label{fn:signature-four}For every $g\geq 3$, there exists a $\Sigma_g^2$-bundle $W^4\to Y^2$ with signature $4$ and a section $s\colon Y^2\to W^4$ of self-intersection zero~\cite[Section~4]{Baykur:2023geography}.
Then if we choose $B=b\,\operatorname{PD}_{W^4}[s(Y^2)]\bmod n$ for $b=\langle\hat{B},[\Sigma_g^2]\rangle\in\mathbb{Z}_n$,
we have $B|_{\Sigma_g^2}=\hat{B}$ and $\int_{W^4}\mathcal{P}(B)=b^2[s(Y^2)]^2=0$.}
Then we have a constraint
\begin{equation}
    N(\Sigma^2,\hat{B})\,\frac{c_-}{24}\int_{W^4}p_1(TW^4) \in \mathbb{Z}.
\end{equation}
Moreover, the bundle $W^4$ in footnote~\ref{fn:signature-four} has signature $4$, and so $\int_{W^4}p_1(TW^4)=3\,\sigma(W^4)=12$. Then we have
\begin{cor}\label{cor:chiral-central-charge}
Consider a $(2+1)$-dimensional TQFT specified by a modular tensor category, with chiral central charge $c_-$. Let $\Sigma_g^2$ be a closed oriented surface of genus $g\geq 3$, and let $\hat{B}$ denote a flux insertion of abelian anyons on $\Sigma_g^2$. The ground-state degeneracy $N(\Sigma_g^2,\hat{B})$ in this background satisfies
\begin{equation}
    N(\Sigma_g^2,\hat{B})\,\frac{c_-}{2} \in \mathbb{Z},\quad g\geq 3.
\end{equation}
This generalizes the result of \cite{Kong:2020bmb} to any flux insertion $\hat{B}$ of abelian anyons.
\end{cor} 
For an insertion of non-abelian anyons, we do not have a similar interpretation of the bulk SPT action. Nevertheless, we expect the same constraints to hold for the ground-state degeneracy $N(\Sigma^2,\hat{B})$ for any flux insertion $\hat{B}$ of non-abelian anyons; insertion of the non-invertible anyon defect line only in the spatial manifold $\Sigma^2$ does not affect the bulk SPT action. Having tested this conjecture numerically for several TQFTs, we have not found any counterexample to this expectation.

When $g\leq 2$, the integral of the Pontryagin class $\int_{W^4}p_1(TW^4)$ is always trivial~\cite{Baykur:2023geography}, and so we have
\begin{equation}
    N(\Sigma_g^2,\hat{B})\times\nu\int_{W^4}\mathcal{P}(B) \in \mathbb{Z}.
\end{equation}
Now take a fibered space $W^4$ to be $\Sigma_{g}^2\times T^2$. For any $g\geq 1$, choose integral one-cycles $a_1,a_2$ in $\Sigma_g^2$ and
$b_1,b_2$ in $T^2$ with oriented intersection numbers
$a_1\cdot a_2=b_1\cdot b_2=1$.
Then define the background gauge field $B$ as
\begin{equation}
    B=\operatorname{PD}(a_1)\cup \operatorname{PD}(b_1)+\operatorname{PD}(a_2)\cup \operatorname{PD}(b_2) \mod n.
\end{equation}
The resulting integral of the Pontryagin square is
\begin{equation}
    \int_{W^4}\mathcal{P}(B)=-2. 
\end{equation}
Now the flux insertion of the spatial manifold $\Sigma_g^2$ is trivial, and so we have a constraint
\begin{equation}
    N(\Sigma_g^2)\times 2\nu \in \mathbb{Z},\quad g\geq 1.
\end{equation}
Now note that the coefficient $\nu$ describes the topological spin of the abelian anyon~\cite{Hsin:2018vcg}. We therefore obtain the following result.
\begin{cor}
    For any abelian anyon $a$ with a topological spin $\theta_a\in \mathbb{R}/\mathbb{Z}$ in a $(2+1)$D TQFT, the ground-state degeneracy $N(\Sigma_g^2)$ of the TQFT on a closed oriented surface $\Sigma_g^2$ of genus $g\geq 1$ satisfies
    \begin{equation}
        N(\Sigma_g^2)\times 2\theta_a \in \mathbb{Z},\quad g\geq 1.
    \end{equation}
\end{cor}
One can actually show this directly: let $a$ be an abelian anyon with topological spin $\theta_a$, and let $U_x,U_y$ be the Wilson loop operators of $a$ acting on $Z(\Sigma_g^2)$ along oriented simple closed curves $x,y$ that intersect transversely at one point with $x\cdot y=1$. Then the commutation relation of these operators is given by
\begin{equation}
    U_xU_y=\exp(4\pi i\theta_a)U_yU_x.
\end{equation}
Taking the determinant of both sides, we have
\begin{equation}
    \det(U_x)\det(U_y)=\exp(4\pi i\theta_aN(\Sigma_g^2))\det(U_y)\det(U_x).
\end{equation}
Since $a$ is invertible, both operators have nonzero determinants. Thus, we obtain the same constraint as above.

\subsection{Applications to symmetry-preserving TQFT}\label{sec:application_sym_pres_tqft}
Stable (simple) unitary TQFTs where a global finite higher-group $\mathbb{G}$-symmetry is not spontaneously broken in the ground states are called \textit{symmetry-preserving TQFTs}. Any symmetry-preserving TQFT has a unique ground state on the sphere $S^{d-1}$ under any $\mathbb{G}$-twisted boundary conditions.%
\footnote{For $d\geq2$, a sphere background, where one cannot activate any nontrivial symmetry defect along spatial directions, is represented by an invertible symmetry line extending in the time direction. If this line were nontrivial, the principle of remote detectability~\cite{Kong:2014qka,Lan:2018vjb} would give a topological $(d-2)$-dimensional operator charged under the symmetry, contradicting symmetry preservation. The line is therefore equivalent to the identity, and simplicity implies that the twisted state space is one-dimensional.}
Then we have
\begin{prop}\label{prop:sym-pres-tqft}
    If an anomaly $\omega\in\operatorname{Tor}\Omega_{\SO}^{d+1}(B\mathbb{G})$ of a finite higher-group $\mathbb{G}$ symmetry admits a symmetry-preserving TQFT, then the following holds:
    \begin{equation}
        \int_{W^{d+1}}g^*\omega=0\pmod{1},
    \end{equation}
    for every bundle $S^{d-1}\rightarrow W^{d+1}\rightarrow Y^2$ over a closed connected surface $Y^2$, with oriented total space $W^{d+1}$, and every classifying map $g\colon W^{d+1}\rightarrow B\mathbb{G}$.
    Here, the integral denotes the $\mathbb{R}/\mathbb{Z}$-valued evaluation of the anomaly on the oriented bordism class of $(W^{d+1},g)$.
\end{prop}

\paragraph{Example.}
Let $G$ be a finite group. Any bosonic anomaly $\omega\in\rH^{d+1}(BG,\U(1))$ of a $0$-form $G$ symmetry admits symmetry-preserving TQFTs when $d>2$ \cite{Witten:2016cio,Wang:2017loc,Tachikawa:2017gyf}. Then we have
\begin{equation}
    \int_{S^{d-1}\rightarrow W^{d+1}\rightarrow Y^2}g^*\omega=0\quad (d>2), 
\end{equation}
where $g\colon W^{d+1}\rightarrow BG$, $\omega\in \rH^{d+1}(BG,\U(1))$. This math statement is rather trivial though, since $g^*\omega$ is pulled back from the two-dimensional base $Y^2$ and hence vanishes for $d>2$.%
\footnote{We can show this as follows. When $d>2$, $\pi_1(S^{d-1})=0$, and so the homotopy exact sequence of the fibration $S^{d-1}\rightarrow W^{d+1}\rightarrow Y^2$ reads
\begin{equation}
    \cdots\rightarrow \pi_2(Y^2)\rightarrow \pi_1(S^{d-1})\rightarrow \pi_1(W^{d+1})\rightarrow \pi_1(Y^2)\rightarrow 0.
\end{equation}
Since $\pi_1(S^{d-1})=0$, we have $\pi_1(W^{d+1})\cong \pi_1(Y^2)$. On the other hand, when $G$ is a finite group, the classifying map $g\colon W^{d+1}\rightarrow BG$ factors through the classifying space of the fundamental group $B\pi_1(W^{d+1})\xrightarrow{\simeq} B\pi_1(Y^2)\rightarrow BG$. Therefore, $g$ factors through $Y^2$ up to homotopy, and hence $g^*\omega=0$.}
On the other hand, one cannot conclude that there is a symmetry-preserving TQFT when $d=1,2$. Indeed, for any nontrivial anomaly of abelian groups, one can always find $W^{d+1}$ such that the evaluation of the anomaly is nontrivial. This is consistent with the fact that there is no topological order in zero and one spatial dimensions.

\paragraph{Anomalous $1$-form symmetry in 3D TQFTs}
Consider a $\mathbb{Z}_p$ $1$-form symmetry with an anomaly of the form
\begin{equation}
    \frac{k}{\gcd(2,p)p}\int\mathcal{P}(B),
\end{equation}
where $k\in\mathbb{Z}_{\gcd(2,p)p}$ and $B$ is the background gauge field for the symmetry.
To rule out symmetry-preserving TQFTs for these anomalies, we can take a nontrivial fibered space $W^4$ of the form $S^2\rightarrow W^4\rightarrow S^2$, which is diffeomorphic to $\mathbb{C}P^2\#\overline{\mathbb{C}P^2}$.
Write $W^4=\mathbb{C}P_1^2\#\overline{\mathbb{C}P_2^2}$, and take $B$ to be the mod-$p$ reduction of the generator of $\rH^2(\mathbb{C}P_1^2,\mathbb{Z})$, i.e., $B$ is the Poincar\'e dual of $\mathbb{C}P_1^1\subset \mathbb{C}P_1^2$ in $W^4$.
Then we have
\begin{equation}
    \frac{k}{\gcd(2,p)p}\int_{W^4}\mathcal{P}(B)=\frac{k}{\gcd(2,p)p}\int_{\mathbb{C}P_1^2}\mathcal{P}(B)=\frac{k}{\gcd(2,p)p}\int_{\mathbb{C}P_1^2}B^2=\frac{k}{\gcd(2,p)p}\mod 1.
\end{equation}
Then by Proposition \ref{prop:sym-pres-tqft}, we see that all nontrivial anomalies of this type do not admit any symmetry-preserving TQFTs.
This bundle construction and the resulting absence of symmetry-preserving TQFTs can be generalized to a $(d-2)$-form $\mathbb{Z}_2$ symmetry with the fermionic anomaly $\omega=\operatorname{Sq}^2(B)/2$ for $d\geq3$.

For the case of $\nu=k/(\gcd(2,p)p)\neq 1/2$, this result is shown in the paper by Cordova--Ohmori \cite{Cordova:2019bsd}. If we take $W^4$ to be a trivial fibered space, say $S^2\times S^2$, we can also apply Proposition \ref{prop:sym-pres-tqft} except for $\nu=1/2$.

\paragraph{Mixed anomaly with $(d-2)$-form symmetry}
Consider a mixed anomaly between a $\mathbb{Z}_n$ $(d-2)$-form symmetry and a $p$-form $A$ symmetry ($p=0,1$) of the form
\begin{equation}
    \frac{k}{\gcd(n,m)}\int_{W^{d+1}}B\cup A^*\widetilde{\omega},
\end{equation}
where $B$ is the background gauge field for the $(d-2)$-form symmetry, and $A^*\omega$ is a nontrivial $2$D SPT action of order $m$ for the $p$-form $A$ symmetry. Precisely, for the class $\omega\in \rH^2(B^{p+1}A,\U(1))$, we choose a lift $\widetilde{\omega}\in \rH^2(B^{p+1}A,\mathbb{Z}_m)$ under the coefficient map $r\mapsto r/m\in\mathbb{R}/\mathbb{Z}$. The cup product $B\cup A^*\widetilde{\omega}$ then takes values in $\rH^{d+1}(W^{d+1},\mathbb{Z}_{\gcd(n,m)})$.
Then the SPT response class is obtained by the coefficient map $\mathbb{Z}_{\gcd(n,m)}\rightarrow \U(1)$.
Such an anomaly is realized, for example, in the $\mathbb Z_n$ gauge theory, where the magnetic and electric symmetries are $\mathbb Z_n$ $(d-2)$-form and $1$-form symmetries, respectively.

Take $W^{d+1}=S^{d-1}\times Y^2$, where $Y^2$ is a generating manifold of the SPT $A^*\omega$, i.e., there is a nontrivial background field such that $\int_{Y^2} A^*\omega =1/m\mod 1$ ($\int_{Y^2}A^*\widetilde{\omega}=1 \mod m$).
Then choose $A$ as such a nontrivial field configuration on $Y^2$, and $B=\operatorname{PD}(\{\mathrm{pt}\}\times Y^2)$.
Then we have
\begin{equation}
    \frac{k}{\gcd(n,m)}\int_{W^{d+1}}B\cup A^*\widetilde{\omega}=\frac{k}{\gcd(n,m)}\mod 1.    
\end{equation}
Thus, all nontrivial mixed anomalies of this type do not admit any symmetry-preserving TQFTs.

\subsection{Mathematical application: fiber bundles as bordism generators}\label{sec:math_application}
Consider a system with a torsion anomaly $\omega$ on a parameter space $X$. The anomaly of the compactified system is given by the pushforward map
\begin{equation}
    \Omega_{\Sigma}=\int_{\Sigma^{d-1}}g^*\omega,
\end{equation}
where $g\colon W^{d+1}\rightarrow X$ is the parameter map. To what extent can we detect the anomaly $\omega$ from the compactified system?
We expect that the pushforward map is \textit{almost} enough to detect the anomaly, i.e., there exists a fibered space $\Sigma^{d-1}\rightarrow W^{d+1}\rightarrow Y^2$ with a parameter map $g$ such that
\begin{equation}
    \int_{W^{d+1}}g^*\omega\neq 0 \mod 1.
\end{equation}
Indeed, if the image of the pushforward map is always trivial, one may construct a TQFT with a unique ground state for any spatial manifold, which seems inconsistent with the often-believed principle---anomaly is the only obstruction to having a unique ground state. 
Of course, the principle is highly nontrivial in general, so we discuss when the principle holds in the following.

First, consider an anomalous $p$-form symmetry ($1\leq p \leq d-2$) and suppose we have a $d$-dimensional QFT in which the symmetry is spontaneously broken and the ground state on the sphere $S^{d-1}$ or the plane $\mathbb{R}^{d-1}$ is unique.
Since the symmetry is spontaneously broken, we may have a nontrivial codimension-$(p+1)$ excitation on the plane, which carries the anomaly of the $p$-form symmetry. The nontriviality of the anomaly can be diagnosed by its (generalized) self-statistics, namely some inconsistency of the phase depending on the global (gauge) symmetry transformation. 
Then one may find a configuration of the background gauge field such that $\int \Omega_{\Sigma}\neq 0\mod 1$. 

In the above discussion, we have used the following assumptions:
\begin{enumerate}
    \item There exists a TQFT with a unique ground state on the sphere $S^{d-1}$, and the anomalous $p$-form symmetry is spontaneously broken in the TQFT.
    \item The $p$-form symmetry anomaly can be detected by the self-statistics of the codimension-$(p+1)$ excitation, namely a phase ambiguity under global symmetry transformations.%
    \footnote{The relation between generalized statistics and anomalies on lattices is discussed in Refs.~\cite{Kobayashi:2024dqj,Xue:2026oxe}. For finite abelian higher-form symmetries, the authors in~\cite{Kobayashi:2024dqj} conjecture that these statistics capture all anomalies classified by group cohomology (Conjecture~IV.2).}
\end{enumerate}
The uniqueness of the ground state on the sphere is needed because otherwise multiple ground-state sectors can absorb the phase ambiguity of the self-statistics. The existence of such TQFTs for general higher-group anomalous symmetries is ensured when they only contain $p$-form symmetries with $1\leq p \leq d-2$ \cite{Kobayashi:2019lep}.%
\footnote{The idea of construction of such TQFTs is as follows. For any higher-group $\mathbb{G}$, one can find a fibration 
\begin{equation}
    \Omega B\mathbb{G}\rightarrow L\mathbb{G}\rightarrow B\mathbb{G},
\end{equation}
where $B$ denotes the classifying space construction, $\Omega X$ is the loop space of $X$, and $L\mathbb{G}$ is the path space. Now the total space $L\mathbb{G}$ is contractible, and any anomaly of $\mathbb{G}$-symmetry is trivialized in the fibered space.
Then one can construct a TQFT with a unique ground state on the sphere $S^{d-1}$ as an $\Omega^2 B\mathbb{G}\simeq \Omega\mathbb{G}$ gauge theory.}
Note that, in general, one needs to choose a spatial manifold $\Sigma^{d-1}$ with nontrivial lower homotopy groups in order to couple to the background gauge field. The ground state on such a spatial manifold is not necessarily unique.
However, since we are considering higher codimensional excitations, the corresponding excitations cannot separate the space and thus the ground state is locally unique.
Therefore, we expect the discussion still holds for general higher-group anomalous symmetries even when considering nontrivial spatial manifolds.

To justify the second assumption, we may need to assume that the anomaly is purely internal, i.e., neither gravitational nor mixed gravitational. If the anomaly mixes with the gravitational anomaly, it may be hard to consider the corresponding excitations or global symmetry transformations. 

For an anomalous finite $0$-form symmetry $G$, a TQFT in which the symmetry is spontaneously broken cannot have a unique ground state on the sphere $S^{d-1}$, since spontaneous breaking of a $0$-form symmetry leads to multiple ground states.
Nevertheless, in spacetime dimension $d>2$, one can construct a symmetry-preserving TQFT with a unique ground state on the sphere $S^{d-1}$~\cite{Witten:2016cio,Wang:2017loc,Tachikawa:2017gyf}.
An explicit construction of such TQFTs is as follows.
For any finite group $G$ and an anomaly $[\omega]\in \rH^{d+1}(BG,\U(1))$ with $d>2$, one can always find a short exact sequence
\begin{equation}
    1\rightarrow A\rightarrow H\xrightarrow{p}  G\rightarrow 1,
\end{equation}
such that $[p^*\omega]=0\in \rH^{d+1}(BH,\U(1))$. Then a symmetry-preserving TQFT is realized as an $A$-gauge theory whose partition function is given by
\begin{equation}
    Z(M^d,g)=\#\sum_{a\in C^1(M^d,A),b\in C^{d-2}(M^d,\hat{A})}\exp\left(2\pi i\int_{M^d}b\cup\delta a+b\cup g^*e+(-1)^{d-1}g^*z\cup a\right),
\end{equation}
where $\hat{A}\coloneqq \operatorname{Hom}(A,\U(1))$ is the Pontryagin dual of $A$ and $e\in Z^2(BG,A),z\in Z^{d-1}(BG,\hat{A})$ are chosen to satisfy $\omega=z\cup e$.%
\footnote{In general, $A$ is a finite abelian $G$-module. The cochains and coboundaries in this formula use the local systems $A_g$ and $\hat{A}_g$ induced by $g$, with cup products paired into $\U(1)$~\cite[Sec.~2.7]{Tachikawa:2017gyf}.}
The factor $\#$ is a normalization factor that depends on the topology of $M^d$.

Note that this theory is just an $A$-gauge theory, and so the two symmetries, a $(d-2)$-form symmetry and a $1$-form symmetry, generated by $\oint_{M^{1}}a$ and $\oint_{M^{d-2}}b$ are spontaneously broken.
Looking at the action of the symmetry-preserving TQFT, one can see that the anomaly is saturated by a symmetry fractionalization between the $G$-symmetry and the two broken symmetries.
Combining this observation with our discussion of the detectability of higher-form anomalies, we expect that one can also find a fibered space with an appropriate configuration of the background gauge field such that the evaluation of the anomaly is nontrivial.

The whole argument above is not rigorous, and we have not given a mathematical proof, though it seems to be physically reasonable. 
Here we just propose the following conjecture:
\begin{conj}
    Let $\mathbb{G}$ be a finite $(d-1)$-group and $\omega\in\operatorname{Tor}\Omega_{\SO}^{d+1}(B\mathbb{G})$ be a nontrivial anomaly of a $\mathbb{G}$-symmetry in spacetime dimension greater than $2$. We further assume that $\omega$ is purely internal, i.e., neither a gravitational nor a mixed gravitational anomaly.
    Then there exists a fibered space $\Sigma^{d-1}\rightarrow W^{d+1}\rightarrow Y^2$ and a classifying map $g:W^{d+1}\rightarrow B\mathbb{G}$ such that
    \begin{equation}
        \int_{W^{d+1}}g^*\omega\neq 0 \mod 1.
    \end{equation}
\end{conj}
We expect that the assumptions in the conjecture cannot simply be dropped.
For instance, there appear to be anomalies of finite $0$-form symmetries in two spacetime dimensions that cannot be detected by any fiber bundle with one-dimensional fibers; by the discussion in Sec.~\ref{sec:application_sym_pres_tqft}, such a symmetry group would necessarily be non-abelian.
Mixed anomalies between a $0$-form symmetry and gravity in four spacetime dimensions may provide similar examples.
We leave a detailed analysis of these cases to future work.

\section{Discussion}\label{sec:discussion}
In this paper, we formulated anomalies of TQFTs as obstructions to constructing parametrized families and derived constraints on ground-state degeneracies and the existence of symmetry-preserving TQFTs.
We conclude with the following open questions.
\begin{itemize}
    \item In this paper, we only considered anomalies of invertible symmetries in bosonic TQFTs.
    Generalization to other theories with different tangential structures, such as spin TQFTs, is a straightforward extension of our work. It is also interesting to consider generalizations to non-invertible symmetries.
    \item The moduli space of parametrized continuum TQFTs, or the target space of the anomaly map, may or may not be equal to that of models on lattices. Studying the agreement or disagreement between these two moduli spaces is an interesting question.
\end{itemize}

\section*{Acknowledgments}
T.A.~is supported by JSPS KAKENHI Grant Number 25KJ1557.
T.A.~was also supported by JST CREST (Grant No.~JPMJCR19T2) and JSPS Overseas Challenge Program for Young Researchers. R.T.~is supported by the Mani L. Bhaumik Presidential Term Chair in Theoretical Physics.
We used OpenAI's ChatGPT (GPT-5.6 Sol and GPT-6 Astra) to discuss mathematical questions and to check the consistency of our arguments.

\appendix
\section{Parametrized TQFTs}\label{app:parametrized-tqft-details}
In this appendix, we provide a more detailed discussion of parametrized TQFTs and their relation to the Postnikov tower of the moduli space of TQFTs.
We mainly focus on the case of non-chiral 3D TQFTs, but most of the discussion can be generalized to higher dimensions.

\subsection{Postnikov tower of Turaev--Viro TQFT}
Turaev--Viro TQFT is a 3D TQFT constructed from a unitary fusion category $\mathcal{C}$~\cite{Turaev:1992hq,Barrett:1993ab}, whose anyonic excitations are described by the Drinfeld center $\mathcal{Z}(\mathcal{C})$~\cite{Turaev:2010pp}.
The corresponding moduli space of parametrized TQFTs is conjectured to be $B\Pic(\mathcal{Z}(\mathcal{C}))\simeq B\BrPic(\mathcal{C})$.
Here, $\BrPic(\mathcal{C})$ denotes the monoidal $2$-groupoid of invertible finite semisimple $\mathcal{C}$-$\mathcal{C}$-bimodule categories~\cite[Sec.~4.1]{Etingof:2009yvg}.
Physically, the first homotopy group $\pi_1(B\Pic(\mathcal{Z}(\mathcal{C})))\cong\operatorname{Pic}(\mathcal{Z}(\mathcal{C}))$ corresponds to the group of invertible domain walls, which is equivalent to the group of isomorphism classes of braided autoequivalences of $\mathcal{Z}(\mathcal{C})$. 
The second homotopy group $\pi_2(B\Pic(\mathcal{Z}(\mathcal{C})))$ describes the group of abelian anyons, which is equivalent to the group of invertible objects in $\mathcal{Z}(\mathcal{C})$.

In the following, we write $\mathcal{B}\coloneqq\mathcal{Z}(\mathcal{C})$. Let us first consider the Postnikov tower of $B\Pic(\mathcal{B})$ from the perspective of deformation classes, which is given by
\begin{equation}
    \begin{tikzcd}
        K(\mathbb{Z},4)
            \arrow[r]
        &
        B\Pic(\mathcal{B})
            \arrow[d]
        \\
        K(\pi_2B\Pic(\mathcal{B}),2)
            \arrow[r]
        &
        \pi_{\leq 2}B\Pic(\mathcal{B})
            \arrow[d]
        \\
        K(\pi_1B\Pic(\mathcal{B}),1)
            \arrow[r]
        &
        \pi_{\leq 1}B\Pic(\mathcal{B})
            \arrow[d]
        \\
        & \ast\mathrlap{.}
    \end{tikzcd}
\end{equation}
Then the $\U(1)$-valued anomaly of an $X$-parametrized family of TQFTs is represented by a map $X\rightarrow K(\mathbb{Z},5)$, whose homotopy class lies in $[X,K(\mathbb{Z},5)]\cong\rH^5(X,\mathbb{Z})$.

\subsection{Obstruction to lifting to \texorpdfstring{$K(\mathbb{Z},4)$}{K(Z,4)}}\label{app:berry-lift}
Consider a parameter family associated with a map $f\colon X\rightarrow\mathcal{M}_{\mathcal{B}}\coloneqq B\Pic(\mathcal{B})$.
The Whitehead tower of $\mathcal{M}_{\mathcal{B}}$ has the following form:
\begin{equation}
    %\begin{tikzcd}
        \ast \xrightarrow{p^4} \mathcal{M}_{\mathcal{B}}^{(3)}\xrightarrow[\simeq]{p^3} \mathcal{M}_{\mathcal{B}}^{(2)} \xrightarrow{p^2} \mathcal{M}_{\mathcal{B}}^{(1)} \xrightarrow{p^1} \mathcal{M}_{\mathcal{B}}^{(0)} \xrightarrow[\simeq]{p^0} \mathcal{M}_{\mathcal{B}}.
    %\end{tikzcd}
\end{equation}
Here, $\mathcal{M}_{\mathcal{B}}^{(k)}$ is $k$-connected ($\pi_{i\leq k}\mathcal{M}_{\mathcal{B}}^{(k)}=0$) and each map $p^k$ has a fiber
\begin{equation}
    K(\pi_k\mathcal{M}_{\mathcal{B}},k-1) \rightarrow \mathcal{M}_{\mathcal{B}}^{(k)} \xrightarrow{p^k} \mathcal{M}_{\mathcal{B}}^{(k-1)},\qquad k\geq 1.
\end{equation}
Thus, the obstruction to having a lift of $f^{k-1}\colon X\rightarrow \mathcal{M}_{\mathcal{B}}^{(k-1)}$ to $\mathcal{M}_{\mathcal{B}}^{(k)}$ is given by
\begin{equation}
    c^{k-1}\circ f^{k-1}\colon X \rightarrow \mathcal{M}_{\mathcal{B}}^{(k-1)}\rightarrow K(\pi_k\mathcal{M}_{\mathcal{B}},k),
\end{equation}
which defines an element of $\rH^k(X,\pi_k\mathcal{M}_{\mathcal{B}})$.

Let us consider a family that only traps anyons, i.e., the truncation map $f_1\colon X\rightarrow \pi_{\leq 1}\mathcal{M}_{\mathcal{B}}$ is trivial.
Let us ask when the parameter dependence of such a family can be realized by a family of invertible states: does $f$ admit a lift to $K(\mathbb{Z},4)$? The relevant stage of the Whitehead tower is as follows:
\begin{equation}
    \settowidth{\simeqKwidth}{${}\simeq K(\mathbb{Z},4)$}
    \begin{tikzcd}
        K(\pi_2\mathcal{M}_{\mathcal{B}},1)
            \arrow[r]
        &
        \mathcal{M}_{\mathcal{B}}^{(2)}\mathrlap{{}\simeq K(\mathbb{Z},4)}
            \arrow[d]
            \arrow[r, shorten <=\simeqKwidth]
        &[\simeqKwidth]
        \ast
            \arrow[d]
        \\
        X
            \arrow[r]
        &
        \mathcal{M}_{\mathcal{B}}^{(1)}
            \arrow[r]
        &
        K(\pi_2\mathcal{M}_{\mathcal{B}},2).
    \end{tikzcd}
\end{equation}
Since $\pi_2\mathcal{M}_{\mathcal{B}}=\mathcal{A}$ is the group of abelian anyons, the obstruction to having a lift to $\mathcal{M}_{\mathcal{B}}^{(2)}\simeq K(\mathbb{Z},4)$ lies in $\rH^2(X,\mathcal{A})$.
For example, when $X=B\U(1)$ the obstruction lies in $\rH^2(B\U(1),\mathcal{A})\cong\mathcal{A}$. Physically, this corresponds to whether the family traps anyons, so one cannot have a lift to $K(\mathbb{Z},4)$ unless the truncation map $f_2\colon X\rightarrow \pi_{\leq 2}\mathcal{M}_{\mathcal{B}}$ is trivial.

\subsection{Fractionalization of higher Berry phase}
Let $X$ be a simply connected space and let us consider a family map $f\colon X\rightarrow B\Pic(\mathcal{B})$.
We can always choose maps $g$ and $s$ as follows:
\begin{equation}
    \begin{tikzcd}
        &
        &
        B\Pic(\mathcal{B})
            \arrow[d,"p_4"]
            \arrow[r]
        & 
        \ast
            \arrow[d]
        \\
        X
            \arrow[rru,"f"]
            \arrow[r,"g"']
        &
        B^2A
            \arrow[r,"s"']
        &
        \pi_{\leq 2}B\Pic(\mathcal{B})
            \arrow[r,"\omega"']
        &
        K(\mathbb{Z},5)\mathrlap{.}
    \end{tikzcd}
\end{equation}
Here, $A\subseteq\pi_2 B\Pic(\mathcal{B})$ is an abelian group of anyons trapped by the map $f$ and $s$ is the inclusion. Due to the anomaly of the $1$-form symmetry $A$, the map $\omega\circ s$ is not trivial in general, and it gives the bulk SPT action represented by a class in $\rH^5(B^2A,\mathbb{Z})$, which is always torsion.
On the other hand, since we have a map $f\colon X\rightarrow B\Pic(\mathcal{B})$, the composed map $\omega\circ s\circ g$ is homotopic to the null map:
\begin{equation}
    \begin{tikzcd}[row sep=large, column sep=large]
        X \vphantom{K(\mathbb{Z},5)}
            \arrow[r, bend left=45, "\omega\circ s \circ g",
            ""{name=U, inner sep=1pt, below}]
            \arrow[r, bend right=45, "\ast"',
            ""{name=D, inner sep=1pt, above}]
        &
        K(\mathbb{Z},5)\mathrlap{.}
            \arrow[Rightarrow, from=U, to=D, shorten=2pt, "\alpha"]
    \end{tikzcd}
\end{equation}
In other words, this homotopy $\alpha\in\operatorname{Map}(X\times[0,1],K(\mathbb{Z},5))$ gives a cocycle $\tilde{\alpha}\in Z^5(X\times[0,1],\mathbb{Z})$.
Then we can obtain a 4-cochain $x$ as
\begin{equation}
    x\coloneqq -\int_{[0,1]}\tilde{\alpha}.
\end{equation}
This gives a trivialization of the anomaly: $\delta x=g^*s^*\omega$.
Now since $s^*\omega$ is torsion, we have $N\cdot s^*\omega=\delta\gamma$ for an integer $N$ and a 4-cochain $\gamma\in C^4(B^2A,\mathbb{Z})$. Then we obtain $g^*\gamma/N\in C^4(X,\mathbb{Q})$ such that $\delta(g^*\gamma/N)=g^*s^*\omega$.
Now
\begin{equation}
    \frac{1}{N}\Omega\coloneqq\frac{g^*\gamma}{N}-x \in Z^4(X,\mathbb{Q})
\end{equation}
describes a fractionalization of the higher Berry phase $[\Omega]\in \rH^4(X,\mathbb{Z})$.

As an example of a parameter family with a nontrivial fractionalization of the higher Berry phase, consider an $S^2\times S^2$-family of Turaev--Viro TQFTs with $\mathcal{C}=\mathrm{Vect}_{\mathbb{Z}_2}$, the category of $\mathbb{Z}_2$-graded vector spaces.
The corresponding Turaev--Viro TQFT is the $\mathbb{Z}_2$ gauge theory, which is realized as the low-energy effective theory of the toric code model.
The anyonic excitations of the $\mathbb{Z}_2$ gauge theory form $\mathbb{Z}_2\times\mathbb{Z}_2$ generated by the $e$ and $m$ anyons.
The action of the $\mathbb{Z}_2$ gauge theory is given by
\begin{equation}
    \frac{1}{2}\int_{M^3} a\cup \delta b,
\end{equation}
where $a,b\in C^1(M^3,\mathbb{Z}_2)$ are dynamical $\mathbb{Z}_2$ gauge fields.
Now consider a parameter space $X=(S^2)_1\times (S^2)_2$ coupled to the $\mathbb{Z}_2$ gauge theory through a map $\sigma\colon M^3\rightarrow X$, with action
\begin{equation}\label{eq:TC_S2S2}
    \frac{1}{2}\int_{M^3} \left(a\cup \delta b+a\cup \sigma^*\mathrm{vol}((S^2)_1)+\sigma^*\mathrm{vol}((S^2)_2)\cup b\right)+\frac{1}{2}\int_{M^3} \sigma^*u_2\cup\sigma^*\mathrm{vol}((S^2)_1).
\end{equation}
Here, $\mathrm{vol}((S^2)_i)$ denotes an integral 2-cocycle normalized by $\int_{(S^2)_i}\mathrm{vol}((S^2)_i)=1$.
The last term is a local counterterm, with $\delta u_2=\mathrm{vol}((S^2)_2)$.
Specifically, $u_2$ is a \v{C}ech cochain given by a smooth real lift of the $\mathbb{R}/\mathbb{Z}$-valued transition functions of the Hopf line bundle on $(S^2)_2$, chosen so that $\delta u_2=\mathrm{vol}((S^2)_2)$.
This family trivializes the pullback of the anomaly of the $A=\mathbb{Z}_2\times\mathbb{Z}_2$ $1$-form symmetry whose bulk SPT action is given by
\begin{equation}
    s^*\omega=\frac{1}{2}C_2\cup C_1\in \rH^4(B^2\mathbb{Z}_2\times B^2\mathbb{Z}_2,\U(1))\cong \rH^5(B^2\mathbb{Z}_2\times B^2\mathbb{Z}_2,\mathbb{Z}).
\end{equation}
The displayed isomorphism is the Bockstein homomorphism.
Indeed, the parameter space is coupled to the $\mathbb{Z}_2$ gauge theory through the map
\begin{equation}
    g\colon (S^2)_1\times (S^2)_2\rightarrow B^2\mathbb{Z}_2\times B^2\mathbb{Z}_2; \quad g^*C_i=\mathrm{vol}((S^2)_i)\bmod 2\quad (i=1,2),
\end{equation}
and the pulled-back anomaly is trivialized because $\rH^5(S^2\times S^2,\mathbb{Z})=0$.
Then the following closed lift of the pulled-back anomaly
\begin{equation}
    \frac{1}{2}\mathrm{vol}((S^2)_2)\cup \mathrm{vol}((S^2)_1)
\end{equation}
provides a nontrivial fractionalization of the higher Berry phase
\begin{equation}
    \mathrm{vol}((S^2)_2)\cup \mathrm{vol}((S^2)_1)\in \rH^4((S^2)_1\times (S^2)_2,\mathbb{Z}).
\end{equation}

\subsection{Isomorphism class perspective}
If we take the discrete topology on the universal $2$-form symmetry $\U(1)_{\mathrm{disc}}$, the Postnikov tower of $B\Pic(\mathcal{B})$ is given by
\begin{equation}
    \begin{tikzcd}
        K(\U(1)_{\mathrm{disc}},3)
            \arrow[r]
        &
        B\Pic(\mathcal{B})
            \arrow[d]
        \\
        K(\pi_2B\Pic(\mathcal{B}),2)
            \arrow[r]
        &
        \pi_{\leq 2}B\Pic(\mathcal{B})
            \arrow[d]
        \\
        K(\pi_1B\Pic(\mathcal{B}),1)
            \arrow[r]
        &
        \pi_{\leq 1}B\Pic(\mathcal{B})
            \arrow[d]
        \\
        & \ast\mathrlap{.}
    \end{tikzcd}
\end{equation}
In this topology, the fractionalization of the higher Berry phase corresponds to an anomaly obstructing a lift to $B\Pic(\mathcal{B})$, as illustrated below:
\begin{equation}
    \begin{tikzcd}
        &
        B\Pic(\mathcal{B})
            \arrow[d,"p_3"]
            \arrow[r]
        & 
        \ast
            \arrow[d]
        \\
        X
            \arrow[ru, dashed, "\nexists\,f"{description}]
            \arrow[r,"g"']
        &
        \pi_{\leq 2}B\Pic(\mathcal{B})
            \arrow[r,"\omega"']
        &
        K(\U(1)_{\mathrm{disc}},4)\mathrlap{.}
    \end{tikzcd}
\end{equation}
The example of the $\mathbb{Z}_2$ gauge theory coupled to $S^2\times S^2$ as in \eqref{eq:TC_S2S2} has a nontrivial anomaly
\begin{equation}
    g^*\omega =\frac{1}{2}\mathrm{vol}((S^2)_2)\cup \mathrm{vol}((S^2)_1)\neq 0\in \rH^4(X,\U(1)_{\mathrm{disc}}).
\end{equation}
Although $2g^*\omega=0$ in $\rH^4(X,\U(1)_{\mathrm{disc}})$, twice the rational representative $\frac{1}{2}\mathrm{vol}((S^2)_2)\cup \mathrm{vol}((S^2)_1)$ of the fractional higher Berry phase gives the nontrivial integral class $\mathrm{vol}((S^2)_2)\cup \mathrm{vol}((S^2)_1)\in\rH^4(X,\mathbb{Z})$.
%We also see an example of fractionalization

\section{Parametrized TQFTs acting on boundary theories}\label{app:bulk-boundary-parametrized-tqft}
Let us consider the $\mathbb{Z}_2$ gauge theory coupled to $S^2$. Let $a,b$ be $2\pi$-periodic $\U(1)$ gauge fields, and normalize $\mathrm{vol}(S^2)$ to have integral one. On a closed $M^3$, its partition function is
\begin{equation}\label{eq:TC_m}
    Z(M^3;\sigma\colon M^3\rightarrow S^2)=\int\mathcal{D}a\mathcal{D}b\,\exp\left(i\frac{2}{2\pi}\int_{M^3}b\wedge da+i\int_{M^3}b\wedge\sigma^*\mathrm{vol}(S^2)\right).
\end{equation}
Put the system on $M^2\times[0,\epsilon]$ with the Dirichlet boundary conditions
$a|_{M^2\times\{0\}}=dc_1$ and $a|_{M^2\times\{\epsilon\}}=dc_2$, where $c_1,c_2$ are compact scalars. Take
\begin{equation}
    \sigma\colon M^2\times [0,\epsilon] \rightarrow S^2;\quad
    (x,x_3)\mapsto \left(f(x),\frac{\pi x_3}{\epsilon}\right),
\end{equation}
where $f\colon M^2\rightarrow S^1$, $\phi$ is the angular coordinate on $S^1$, and the second coordinate is the polar angle $\theta$ on $S^2$. Choose the orientation of $S^2$ so that $\int_{[0,\epsilon]}\sigma^*\mathrm{vol}(S^2)=f^*(d\phi/(2\pi))$. Keeping the relative holonomy, we integrate out the remaining interior modes of $a$ and obtain
\begin{equation}
    \frac{2}{2\pi}\int_{M^2}\left(dc_2-dc_1-d\int_0^\epsilon a_3\,dx_3\right)\wedge b
    +\int_{M^2}b\wedge f^*\left(\frac{d\phi}{2\pi}\right).
\end{equation}
Writing $c=c_2-c_1-\int_0^\epsilon a_3\,dx_3\bmod 2\pi$ for the gauge-invariant compact scalar and $b$ for the descended connection on $M^2$, we obtain
\begin{equation}
    \frac{2}{2\pi}\int_{M^2}dc\wedge b+\int_{M^2}b\wedge f^*\left(\frac{d\phi}{2\pi}\right).
\end{equation}
This action is the $(1+1)$D $\mathbb{Z}_2$ gauge theory with a nontrivial coupling $f\colon M^2\rightarrow S^1$.

This result can also be seen in the ``textured'' picture: $S^2$-parameter families correspond to codimension-two invertible defects, represented here by abelian anyon lines. The coupling term in \eqref{eq:TC_m} traps a magnetic $m$-anyon, which does not condense at the Dirichlet boundary. Shrinking the 3D bulk leaves a nontrivial Wilson line in the 2D theory, a codimension-one defect in its spacetime. Thus the resulting $\mathbb{Z}_2$ gauge theory forms a nontrivial $S^1$-parameter family.
This construction produces a 2D parametrized theory from a 3D bulk family and a choice of boundary condition. Changing the boundary condition may produce another 2D family. Hence the bulk-boundary correspondence can be phrased as \textit{bulk families act on boundary families}, as we discuss below.

\paragraph{Categorical perspective}
Here we consider $(2+1)$D TQFTs that admit gapped boundary conditions.
Such a TQFT is known to be equivalent to a Turaev--Viro type TQFT \cite{Fuchs:2012dt}. The moduli space of the Turaev--Viro theory associated with a fusion category $\mathcal{C}$ may be $B\BrPic(\mathcal{C})\simeq B\mathrm{Mod}(\mathcal{Z}(\mathcal{C}))^{\mathrm{inv}}\simeq B\mathrm{Bimod}(\mathcal{C})^{\mathrm{inv}}$~\cite[Prop.~3.11]{Etingof:2009yvg}.
Here, $\mathrm{Mod}(\mathcal{C})$ denotes the $2$-category of finite semisimple $\mathcal{C}$-module categories, with module functors as $1$-morphisms and module natural transformations as $2$-morphisms. Similarly, $\mathrm{Bimod}(\mathcal{C})$ denotes the $2$-category of finite semisimple $\mathcal{C}$-$\mathcal{C}$-bimodule categories.
The superscript $\mathrm{inv}$ retains objects invertible under the relative tensor product, equivalences, and invertible $2$-morphisms.
We use bulk families parametrized by a based topological space $\Sigma Y$, where $\Sigma$ is the reduced suspension.
We have
\begin{equation}
    [\Sigma Y,B\mathrm{Bimod}(\mathcal{C})^{\mathrm{inv}}]_{\ast}\cong[Y,\Omega B\mathrm{Bimod}(\mathcal{C})^{\mathrm{inv}}]_{\ast}
    \cong[Y,\mathrm{Bimod}(\mathcal{C})^{\mathrm{inv}}]_{\ast}
    \cong[Y,\operatorname{End}(\mathrm{Mod}(\mathcal{C}))^{\mathrm{inv}}]_{\ast}.
\end{equation}
Here, $\operatorname{End}(\mathrm{Mod}(\mathcal{C}))$ denotes the monoidal $2$-category of linear $2$-endofunctors of $\mathrm{Mod}(\mathcal{C})$, with composition as its tensor product.
The first identification follows from the suspension--loop adjunction, and the last from the categorical Eilenberg--Watts theorem~\cite[Thm.~3.2.2 (Thm.~2.2.2 in the arXiv version)]{Decoppet2022}.
Thus, bulk parameter families define actions on boundary families classified by 
\begin{equation}
    [Y,\mathrm{Mod}(\mathcal{C})^{\simeq}]_{\ast}.
\end{equation}
Here, $\mathrm{Mod}(\mathcal{C})^{\simeq}$ denotes the maximal $2$-groupoid of $\mathrm{Mod}(\mathcal{C})$, with all objects, module equivalences as $1$-morphisms, and invertible natural transformations as $2$-morphisms. In the expression above, we use its classifying space.
Suppose $Y$ is connected and fix an indecomposable $\mathcal{C}$-module category $M$ at the base point.
The corresponding connected component of the boundary moduli space is $B(\mathcal{C}^M)^{\mathrm{inv}}$, where $\mathcal{C}^M\coloneqq\operatorname{Fun}_{\mathcal{C}}(M,M)$ is the fusion category of module endofunctors. Let $G_M$ be the group of isomorphism classes of invertible objects of $\mathcal{C}^M$. The Postnikov tower of this connected component is given by the following fibration:
\begin{equation}
    \begin{tikzcd}
        B^2\U(1) \arrow[r] & B(\mathcal{C}^M)^{\mathrm{inv}} \arrow[d]\\
        & K(G_M,1)\mathrlap{.}
    \end{tikzcd}
\end{equation}
Now the homotopy type of the fiber $B^2\U(1)$ again depends on the choice of topology on $\U(1)$.

\section{Anomaly of parametrized TQFTs with global symmetry}\label{app:anomaly_with_symmetry}
In this Appendix,%
\footnote{We received substantial assistance from ChatGPT (GPT-6 Astra) in preparing this appendix.}
we discuss moduli spaces and anomalies of parametrized TQFTs with global symmetries.
Throughout this appendix, we work over $\mathbb{C}$.
Let us start with a general $d$-dimensional TQFT with a unitary fusion $(d-1)$-category $\mathcal{C}$ symmetry, which may be classified by a $\mathcal{C}$-module $(d-1)$-category $M$.
We assume that $M$ is finite semisimple.
We note that $M$ can be decomposed into simple TQFTs if we forget the symmetry. Stable $\mathcal{C}$-symmetric TQFTs are classified by indecomposable $\mathcal{C}$-module categories.%
\footnote{The classification in $d=2$ was established in~\cite[Sec.~2.4]{Thorngren:2019iar} and~\cite[App.~A.3]{Komargodski:2020mxz}.}
The self-interfaces of a general $\mathcal{C}$-symmetric TQFT $M$ are classified by $\mathcal{C}^M\coloneqq\operatorname{Fun}_{\mathcal{C}}(M,M)$~\cite{Inamura:2024jke}.
Here, $\operatorname{Fun}_{\mathcal{C}}(M,M)$ is the monoidal $(d-1)$-category of $\mathcal{C}$-module endofunctors, with composition as its tensor product.
In particular, invertible self-interfaces are classified by the group of isomorphism classes of invertible objects of $\mathcal{C}^M$, which we denote by $G_M$.
For a $\mathcal{C}$-module category $M$,
%such that compatible with the module structure everywhere in parameter spaces, 
%then there are no chance to trap nontrivial point operators. Indeed, any nontrivial topological point operator in the TQFT must be charged by a symmetry operators in $\mathcal{C}$.
the moduli space of TQFTs fully compatible with its module structure would be $B\operatorname{Aut}_{\mathcal{C}}(M)\coloneqq B(\mathcal{C}^M)^{\mathrm{inv}}$, as conjectured in \cite{Inamura:2024jke}.
Here, $\operatorname{Aut}_{\mathcal{C}}(M)$ is the monoidal $(d-1)$-groupoid consisting of $\mathcal{C}$-module autoequivalences and invertible higher morphisms.

\subsection{Moduli space without symmetry}
To consider an anomaly of such families, let us consider families of the same TQFT $M$ but without the symmetry.
Since such a TQFT need not be simple after forgetting the symmetry, we need to consider $(d-1)$-form symmetries generated by point operators.
In a $\mathcal{C}$-symmetric TQFT $M$, the algebra of topological point operators contains projectors onto the individual vacuum sectors.
Independent $\U(1)$ phase rotations of these sectors form $\U(1)^N$, where $N$ is the number of vacuum sectors.
This $\U(1)^N$ fits into
\begin{equation}
    1\rightarrow \U(1)^{\mathrm{diag}} \rightarrow \U(1)^N \rightarrow \U(1)^{N-1} \rightarrow 1,
\end{equation}
where $\U(1)^{\mathrm{diag}}$ is the diagonal subgroup of $\U(1)^N$.
Let $\mathcal{M}(M)$ denote the moduli space of the underlying TQFT without the symmetry.
The $\U(1)$-valued anomaly of a projective family $X\rightarrow P\mathcal{M}(M)$ is the obstruction to lifting it to $\mathcal{M}(M)$.
Here, the projective moduli space fits into the fibration
\begin{equation}
    B^d\U(1)^{\mathrm{diag}}\rightarrow \mathcal{M}(M)\rightarrow P\mathcal{M}(M).
\end{equation}
\paragraph*{(1+1)D example}
As an example, consider a two-fold degenerate TQFT in $(1+1)$D, which is equivalent to the $\mathbb{Z}_2$ gauge theory. The moduli space of such TQFTs is $B\operatorname{Aut}_{\mathrm{Vect}}(\mathrm{Vect}^{\oplus 2})\simeq B^2\U(1)^2\mathord{\sslash} \mathbb{Z}_2$, which has a fibration
\begin{equation}
    K(\mathbb{Z}^2,3) \rightarrow B^2\U(1)^2\mathord{\sslash} \mathbb{Z}_2 \rightarrow B\mathbb{Z}_2.
\end{equation}
Here, we take the standard topology on $\U(1)$. 
$\pi_1B\operatorname{Aut}_{\mathrm{Vect}}(\mathrm{Vect}^{\oplus 2})\cong \mathbb{Z}_2$ is generated by the $\mathbb{Z}_2$ $0$-form symmetry switching the two vacuum sectors, and so
the $\mathbb{Z}_2$ action is given by the permutation of two $\U(1)$ factors.
Take a parameter space $X=B\mathbb{Z}_2$ and fix an isomorphism $\rho\colon\pi_1X\rightarrow\mathbb{Z}_2$.
The homotopy classes of lifts of the corresponding map $X\rightarrow B\mathbb{Z}_2$ to $PB\operatorname{Aut}_{\mathrm{Vect}}(\mathrm{Vect}^{\oplus 2})$ form a torsor for $\rH^3(B\mathbb{Z}_2,(\mathbb{Z}^2/\mathbb{Z}^{\mathrm{diag}})_\rho)\cong \mathbb{Z}_2$.
One of the two choices can be lifted to $B\operatorname{Aut}_{\mathrm{Vect}}(\mathrm{Vect}^{\oplus 2})$, while the other cannot. The corresponding obstruction takes values in $\rH^4(B\mathbb{Z}_2,\mathbb{Z}^{\mathrm{diag}})\cong \mathbb{Z}_2$.
The choice with nonzero obstruction is described by the following action of the $\mathbb{Z}_2$ gauge theory coupled to $B\mathbb{Z}_2$:
\begin{equation}
    \frac{1}{2}\int_{M^2}\left(a\cup \delta\phi+a\cup A+\frac{\delta\tilde{A}}{2}\cup\phi\right).
\end{equation}
Here, $a\in C^1(M^2,\mathbb{Z}_2),\phi\in C^0(M^2,\mathbb{Z}_2)$ are dynamical $\mathbb{Z}_2$ gauge fields, $A\in Z^1(M^2,\mathbb{Z}_2)$ is a background $\mathbb{Z}_2$ gauge field, and $\tilde{A}\in C^1(M^2,\mathbb{Z})$ is an integral lift of $A$. 

More generally, the moduli space of $N$-fold degenerate $(1+1)$D TQFTs is given by $\mathcal{M}=B\operatorname{Aut}_{\mathrm{Vect}}(\mathrm{Vect}^{\oplus N})$, and it fits into the following fibration:
\begin{equation}
    \begin{tikzcd}
        K(\mathbb{Z}^{\mathrm{diag}},3)   \arrow[r] 
        & 
        B\operatorname{Aut}_{\mathrm{Vect}}(\mathrm{Vect}^{\oplus N})
            \arrow[d]
            \arrow[r]
        &
        *
            \arrow[d]
        \\
        & 
        PB\operatorname{Aut}_{\mathrm{Vect}}(\mathrm{Vect}^{\oplus N})
            \arrow[r]
        &
        K(\mathbb{Z}^{\mathrm{diag}},4)\mathrlap{.}
    \end{tikzcd}
\end{equation}
We note that here we consider unitary $(1+1)$D TQFTs up to deformation, allowing the positive Frobenius trace weights of the vacuum sectors to vary continuously~\cite[Sec.~3.1]{Moore:2006dw}.
Thus, all permutations of the $N$ vacuum sectors are included and we have $\pi_1\mathcal{M}\cong S_N$.

\subsection{Obstruction to symmetric family}
Consider a $d$-dimensional TQFT $M$ with a unitary fusion $(d-1)$-category symmetry $\mathcal{C}$.
Here we give $\U(1)$ the discrete topology and denote the corresponding moduli spaces with and without the symmetry by $\mathcal{M}_{\mathcal{C}}$ and $\mathcal{M}$, respectively.
Since any parametrized family of TQFTs with a globally defined $\mathcal{C}$ symmetry is also a family of TQFTs without the symmetry, we have a map
\begin{equation}
    F\colon \mathcal{M}_{\mathcal{C}}\rightarrow \mathcal{M}.
\end{equation}
We work in the connected components containing the chosen $\mathcal{C}$-module $M$ and its underlying TQFT.
Projectivizing both spaces by the same universal diagonal $\U(1)$ induces $PF\colon P\mathcal{M}_{\mathcal{C}}\rightarrow P\mathcal{M}$.
Fix a projective family $f\colon X\rightarrow P\mathcal{M}$ whose image in $P\mathcal{M}_{\mathrm{std}}$ has a chosen lift $f_{\mathrm{std}}\colon X\rightarrow\mathcal{M}_{\mathrm{std}}$ with the standard topology.
Thus, $f$ may have a $\U(1)$-valued scalar anomaly in the discrete topology, but it is trivialized in the standard topology.
For this fixed $f$, a compatible projective symmetry structure is a lift $\tilde{f}\colon X\rightarrow P\mathcal{M}_{\mathcal{C}}$ together with a homotopy $PF\circ\tilde{f}\simeq f$.
We write $\mathcal{F}_M=\operatorname{hofib}_M(PF)$.
To construct a lift, we use the relative (Moore--Postnikov) tower of $PF$~\cite[Sec.~4.3]{Hatcher2002}; see also~\cite{Robinson:1972moore,Hill:1976moore}:
\begin{equation}
    P\mathcal{M}_{\mathcal{C}}\longrightarrow\cdots\longrightarrow
    (P\mathcal{M}_{\mathcal{C}})_{k+1}\xrightarrow{p_{k+1}}
    (P\mathcal{M}_{\mathcal{C}})_k\longrightarrow\cdots\longrightarrow
    (P\mathcal{M}_{\mathcal{C}})_0\longrightarrow P\mathcal{M}.
\end{equation}
Here, $(P\mathcal{M}_{\mathcal{C}})_k\coloneqq\pi_{\leq k}^{/P\mathcal{M}}(P\mathcal{M}_{\mathcal{C}})$ truncates each fiber while keeping the base $P\mathcal{M}$ fixed.
Thus, we have the homotopy fiber sequences
\begin{equation}
    \begin{aligned}
        \pi_{\leq k}\mathcal{F}_M&\longrightarrow(P\mathcal{M}_{\mathcal{C}})_k\longrightarrow P\mathcal{M},\\
        K(\pi_{k+1}\mathcal{F}_M,k+1)&\longrightarrow(P\mathcal{M}_{\mathcal{C}})_{k+1}\xrightarrow{p_{k+1}}(P\mathcal{M}_{\mathcal{C}})_k.
    \end{aligned}
\end{equation}
For $k\geq1$, the second fibration has a relative Postnikov class $\omega_k\in\rH^{k+2}((P\mathcal{M}_{\mathcal{C}})_k,(\pi_{k+1}\mathcal{F}_M)_{\rho_k})$, where the subscript denotes the induced action $\rho_k$ of $\pi_1((P\mathcal{M}_{\mathcal{C}})_k)$ on $\pi_{k+1}\mathcal{F}_M$.
For a partial lift $f_k\colon X\rightarrow(P\mathcal{M}_{\mathcal{C}})_k$ over $P\mathcal{M}$, the obstruction to lifting it one stage further is
\begin{equation}\label{eq:relative-symmetry-obstruction}
    o_{k+2}(f_k)=f_k^*\omega_k\in\rH^{k+2}(X,(\pi_{k+1}\mathcal{F}_M)_{\rho_k\circ(f_k)_*}),
\end{equation}
where the coefficient action on $X$ is obtained by composing $\rho_k$ with $(f_k)_*\colon\pi_1X\rightarrow\pi_1((P\mathcal{M}_{\mathcal{C}})_k)$.
The same formula applies for $k=0$ when $\pi_1\mathcal{F}_M$ is abelian.
The initial stages involve $\pi_0\mathcal{F}_M$ and the possibly non-abelian group $\pi_1\mathcal{F}_M$.
A lift exists when the initial lifting problems can be solved and all subsequent obstructions vanish for compatible choices of lifts. Higher obstructions can depend on the lower lifts.

The forgetful and projectivization maps fit into the commutative diagram
\begin{equation}
    \begin{tikzcd}
        \mathcal{M}_{\mathcal{C}} \arrow[r,"F"] \arrow[d]
        & \mathcal{M} \arrow[d] \\
        P\mathcal{M}_{\mathcal{C}} \arrow[r,"PF"]
        & P\mathcal{M}\mathrlap{.}
    \end{tikzcd}
\end{equation}
Suppose we are given a projective symmetric family $X\to P\mathcal{M}_{\mathcal{C}}$ and a genuine underlying family $X\to\mathcal{M}$, with a specified homotopy identifying their images in $P\mathcal{M}$.
Since $F$ acts identically on the universal scalar subgroup, we can find a compatible lift $X\to\mathcal{M}_{\mathcal{C}}$, which gives a genuine symmetric family.
This argument applies to either choice of topology on $\U(1)$ and also gives $\mathcal{F}_M\simeq\operatorname{hofib}_M(F)$.
In particular, a projective symmetric lift of $f$ in the discrete topology can be combined with $f_{\mathrm{std}}$ to obtain a genuine symmetric family in the standard topology.

\paragraph*{Obstructions in (1+1)D}
Let $M$ have $N$ simple objects and decompose into $r$ indecomposable $\mathcal{C}$-module categories.
Since $\operatorname{hofib}(PF)\simeq\operatorname{hofib}(F)$, we compute $\mathcal{F}_M$ using the forgetful map before projectivization, with the discrete topology on $\U(1)$:
\begin{equation}
    F\colon B\operatorname{Aut}_{\mathcal{C}}(M)\longrightarrow B\operatorname{Aut}_{\mathrm{Vect}}(M)
    \simeq K(\U(1)_{\mathrm{disc}}^N,2)\mathord{\sslash} S_N.
\end{equation}
It induces a homomorphism $p\colon G_M\rightarrow S_N$ on $\pi_1$ and an injection $i\colon\U(1)_{\mathrm{disc}}^r\hookrightarrow\U(1)_{\mathrm{disc}}^N$ on $\pi_2$.
The latter assigns a common phase to all vacuum sectors in each indecomposable module component.
Let $Q\coloneqq\U(1)_{\mathrm{disc}}^N/i(\U(1)_{\mathrm{disc}}^r)$. The homotopy exact sequence gives
\begin{equation}
    \begin{gathered}
        \pi_0\mathcal{F}_M\cong S_N/\operatorname{im}p,
        \qquad \pi_j\mathcal{F}_M=0\quad(j\geq2),\\
        1\longrightarrow Q\longrightarrow\pi_1\mathcal{F}_M
        \longrightarrow\ker p\longrightarrow1.
    \end{gathered}
\end{equation}
Here, the homotopy groups are taken at the chosen symmetry structure, and $\pi_0\mathcal{F}_M$ is a set of cosets.
The group extension need not split.
The first stages require the vacuum permutation $\rho\colon\pi_1X\rightarrow S_N$ to lift to $\tilde{\rho}\colon\pi_1X\rightarrow G_M$.
With this lift fixed, the remaining lifting problem has fiber $K(Q,1)$ and hence a single obstruction $o_2(f,\tilde{\rho})\in\rH^2(X,Q_{\tilde{\rho}})$.
The coefficient action is induced by the vacuum permutation $p\circ\tilde{\rho}=\rho$, while the obstruction also depends on the coherence data of $PF$.
In particular, $Q\cong\U(1)_{\mathrm{disc}}^{N-r}$, and for indecomposable $M$ it is $\U(1)_{\mathrm{disc}}^N/\U(1)_{\mathrm{disc}}^{\mathrm{diag}}$.

More explicitly, the obstruction class $o_2(f,\tilde{\rho})$ is computed as follows. For simplicity, we consider the case where $M$ is an indecomposable $\mathcal{C}$-module category.
Take a good cover $\{U_i\}$ of $X$. On each intersection $U_i \cap U_j$, we have exchanges of ground states specified by $\{\sigma_{ij}\}$. We have the corresponding functors $P_{ij}$ representing the permutations of the vacuum sectors. Then, by the consistency condition, we have a natural isomorphism $\mu_{ijk}\colon P_{ij}P_{jk}\Rightarrow P_{ik}$ on triple overlaps $U_i\cap U_j\cap U_k$. Since each vacuum is specified up to a phase, the natural isomorphism is specified by a diagonal matrix
\begin{equation}
    \mu_{ijk}=\operatorname{diag}(e^{2\pi i\tilde{b}_{ijk}^{1}},\ldots,e^{2\pi i\tilde{b}_{ijk}^{N}}).
\end{equation}
From this, we obtain a twisted \v{C}ech cocycle $b=(b_{ijk})$ with coefficients in $Q_{\tilde{\rho}}$, satisfying the twisted cocycle condition
\begin{equation}
    \sigma_{ij}\cdot b_{jkl}-b_{ikl}+b_{ijl}-b_{ijk}=0.
\end{equation}

On the other hand, fix a lift $\tilde{\rho}\colon\pi_1X\rightarrow G_M=\pi_0\operatorname{Aut}_{\mathcal{C}}(M)$ of $\rho$, represented by transition functions $g_{ij}$ satisfying $g_{ij}g_{jk}=g_{ik}$ and $p(g_{ij})=\sigma_{ij}$.
We choose $\mathcal{C}$-module natural isomorphisms $\mu_{g,h}^{\mathcal{C}}\colon F_gF_h\Rightarrow F_{gh}$ for $g,h\in G_M$, where $F_g$ is a $\mathcal{C}$-module autoequivalence representing $g$.
After applying the forgetful functor $F$ and identifying the underlying functors with the corresponding strict permutation functors, we can again represent $\mu_{g,h}^{\mathcal{C}}$ by a diagonal matrix
\begin{equation}
    \mu^{\mathcal{C}}(g,h)=\operatorname{diag}(e^{2\pi i\tilde{t}^{1}(g,h)},\ldots,e^{2\pi i\tilde{t}^{N}(g,h)}).
\end{equation}
Denote the corresponding twisted group cocycle by $t\in Z^2(G_M,Q_p)$, where $G_M$ acts on $Q$ through $p$. Pulling back along $\tilde{\rho}$, we obtain a twisted \v{C}ech cocycle with components
\begin{equation}
    (\tilde{\rho}^*t)_{ijk}=t(g_{ij},g_{jk}).
\end{equation}
Extending this lift to a symmetric projective family requires that the two cocycles $b$ and $\tilde{\rho}^*t$ agree up to a coboundary.
The obstruction is then
\begin{equation}\label{eq:symmetric-family-o2}
    o_2(f,\tilde{\rho})=\bigl[\{b_{ijk}-t(g_{ij},g_{jk})\}\bigr]
    =[b-\tilde{\rho}^*t]\in\rH^2(X,Q_{\tilde{\rho}}).
\end{equation}

\paragraph*{Examples}
Let us consider an example where $M=\mathrm{Vect}^{\oplus 2}$, which is realized by the $\mathbb{Z}_2$ gauge theory. Take a parameter space $X=S^2$ coupled to the $\mathbb{Z}_2$ gauge theory as
\begin{equation}
    \frac{1}{2}\int_{M^2}\left(a\cup\delta\phi+g^*\mathrm{vol}(S^2)\cup \phi\right).
\end{equation}
Here, $g\colon M^2\to S^2$ is the parameter map, and $\int_{S^2}\mathrm{vol}(S^2)=1$.
Physically, the parameter $S^2$ traps the nontrivial point operator. This $S^2$ family is nontrivial in the discrete topology, while trivial in the standard topology.
Now consider a symmetry $\mathcal{C}=\mathrm{Vect}_{\mathbb{Z}_2}$. Then the $\mathbb{Z}_2$ gauge theory corresponds to the choice of the regular module $M=\mathcal{C}$. One sees that the $S^2$ family cannot be lifted to a $\mathcal{C}$-symmetric family, since the nontrivial point operator is charged under the $\mathbb{Z}_2$ symmetry generated by the Wilson line.
Explicitly, $b=\mathrm{vol}(S^2)/2$, identifying $Q\cong\U(1)_{\mathrm{disc}}$ by the relative phase of the two vacua. Since $t$ is trivial, \eqref{eq:symmetric-family-o2} gives $o_2(f,\tilde{\rho})=[b]\neq 0$.

\bigskip
As a concrete example where $\tilde{\rho}^*t$ is nontrivial, consider $\mathcal{C}=\mathrm{Vect}_{\mathbb{Z}_2}^{\omega}$, where $\omega\in\rH^3(B\mathbb{Z}_2,\U(1))$ is the nontrivial anomaly. Take $M=\mathrm{Vect}_{\mathbb{Z}_2}$ and $X=\mathbb{R}P^3$. We consider a parameter family of the $\mathbb{Z}_2$ gauge theory with an action
\begin{equation}
    \frac{1}{2}\int_{M^2}\left(a\cup\delta\phi+a\cup g^*A\right),
\end{equation}
where $g\colon M^2\rightarrow\mathbb{R}P^3$ is a map to the parameter space, and $[A]\in \rH^1(\mathbb{R}P^3,\mathbb{Z}_2)$ is the nonzero generator. We note that if we couple the theory to the background $\mathbb{Z}_2$ gauge field $B$ for the $\mathcal{C}=\mathrm{Vect}_{\mathbb{Z}_2}^{\omega}$ symmetry, the action would be
\begin{equation}
    \frac{1}{2}\int_{M^2}\left(a\cup\delta\phi+a\cup (g^*A+B)
    +\frac{\delta\tilde{B}}{2}\cup\phi
     \right),
\end{equation}
where $\tilde{B}\in C^1(M^2,\mathbb{Z})$ is an integral lift of $B$. 
This parameter family has a nontrivial class $[\tilde{\rho}^*t]=[A\cup A/2]\in\rH^2(\mathbb{R}P^3,Q_{\tilde{\rho}})$, while $[b]$ is trivial; hence $o_2(f,\tilde{\rho})\neq0$ and this family admits no compatible $\mathcal{C}$-symmetric structure.

If we instead consider the parameter family
\begin{equation}
    \frac{1}{2}\int_{M^2}\left(a\cup\delta\phi+a\cup g^*A+\frac{\delta(g^*\tilde{A})}{2}\cup\phi\right),
\end{equation}
it has $[b]=[\tilde{\rho}^*t]\neq 0$, and so the obstruction class is trivial. Here, $\tilde{A}\in C^1(\mathbb{R}P^3,\mathbb{Z})$ is an integral lift of $A$.
We note that in the second case, the parameter family has a $\U(1)$-valued anomaly that is nontrivial in the discrete topology but trivial in the standard topology.

\bibliographystyle{taamsalpha}
\bibliography{ref}

\end{document}